\documentclass[twocolumn]{aastex631}

\usepackage{amsmath}
\usepackage{amssymb}
\usepackage{comment}
\usepackage{booktabs}
\usepackage{longtable}

\newcommand{\m}{M87*}
\newcommand{\s}{Sgr\,A*}

\begin{document}

\title{Photon ring polarimetry with next-generation black hole imaging I. M87*}

\author[0000-0001-8763-4169]{Aditya Tamar}
\affiliation{Department of Physics, National Institute of Technology \\
Surathkal, Karnataka-575025, India}

\author[0000-0002-7179-3816]{Daniel~C.~M.~Palumbo}
\affil{Center for Astrophysics $\vert$ Harvard \& Smithsonian, 60 Garden Street, Cambridge, MA 02138, USA}
\affiliation{Black Hole Initiative at Harvard University, 20 Garden Street, Cambridge, MA 02138, USA}



\begin{abstract}
The near-horizon region of a black hole impacts linear (LP) and circular polarization (CP) through strong lensing of photons, adding large-scale symmetries and anti-symmetries to the polarized image. To probe the signature of lensing in polarimetry, we utilise a geometric model of concentric, Gaussian rings of equal radius to investigate the transition in the Fourier plane at which the photon ring signal begins to dominate over the direct image. We find analytic, closed-form expressions for the transition radii in total intensity, LP, and CP, wherein the resultant formulae are composed of ratios of tunable image parameters, with the overall ``scale" set primarily by the thickness of the direct image. Using these formulae, we compute the transition radii for time-averaged images of M87* simulations at 230 GHz, studying both Magnetically Arrested Disc (MAD) and Standard and Normal Evolution (SANE) configurations for various spin and electron heating models. We compare geometric values to radii obtained directly from the simulations through a coherent averaging scheme. We find that nearly all MAD models have a photon ring-dominated CP signal on long baselines shorter than the Earth diameter at 230 GHz. Across favored models for the M87* accretion flow identified by EHT polarimetric constraints, we quantify the sensitivity and antenna size  requirements for the next-generation EHT and the Black Hole Explorer orbiter to detect these features. We find that the stringent requirements for CP favour explorations using long baselines on the ground, while LP remains promising on Earth-space baselines.
\end{abstract}


\section{Introduction}

The work by the Event Horizon Telescope (EHT) collaboration in imaging the horizon scale emission of black holes Messier 87* (hereafter \m) \citep{EHT_M87_I,EHT_M87_IV,EHT_2024} and Sagittarius A* (hereafter \s) \citep{EHT_Sgr_I} using very-long-baseline interferometry (VLBI) has opened up a new window in probing the astrophysical features of the strong field gravity regime. The observation of polarised structure in these near-horizon images \citep{EHT_M87_VII,Roelofs:2023,EHT_M87_IX,EHT_SgrA_VII} allows us to study the behaviour of magnetic fields in the vicinity of black holes \citep{EHT_M87_VIII,Tsunetoe:2020,EHT_SgrA_VIII}. Indeed, the Astro2020 Decadal Survey emphasized the role of the EHT in this regime, noting the importance of probing the base of the relativistic jet of \m, while also recognising the role of such observations to inform our understanding energy extraction from black holes \citep{astro2023}. More recently, images of the base of the \m{} jet at 86 GHz \citep{Lu_2023} have shown a bright ring of light around a flux depression, which taken with EHT results suggest very optically thin emission with rich prospects for accessing gravity signatures near the horizon, in polarization and otherwise.

In the context of future extensions of the EHT, polarimetry has also served as a useful tool for probing the so-called ``$n=1$ photon ring'' \citep{Johnson_2020,Gralla:2020_grt,Wong:2022,Palumbo:2023}, a strong lensing feature of photons orbiting the black hole with an appearance that is only weakly dependent on the astrophysical details of the accretion.
The ``order" $n$ indexes the photon ring by the number of half-orbits made by the photon (in the $\theta$ direction of the Boyer-Lindquist co-ordinate system) \citep{Gralla:2020,Johnson_2020}. Here the $n=0$ is the image arising from the ``direct" emission and $n=1$ is the secondary image arising from strong lensing.  For an equivalent indexing scheme that is covariant but only works when performing ``forward" ray tracing from the emitting source to the observer, see Appendix A.3 of \citet{Chang:2024}. 
This indexing has been used to probe photon ring signatures in general relativistic magnetohydrodynamic (GRMHD) simulations, \citep{Ricarte:2021,Palumbo:2022,Palumbo:2023}, EHT observations \citep{Broderick:2022} and for analytic modelling of photon ring structures \citep{Tiede:2022,Avendano:2023A,Avendano:2023}.

Through both analytic work and GRMHD simulations, there have been several studies of linear (LP) \citep{Himwich:2020,Rosales:2021,Palumbo:2022,Palumbo:2023} and circular polarization (CP)\citep{Ricarte:2021,MM_2021} features as proxies for inferring the presence of the photon ring, also leading to a rich interplay with polarimetric features in jets \citep{MM:2017,Davelaar:2019,Kawashima:2021,Tsunetoe:2022,Ogihara:2024}.

Now, the EHT observations are performed using only Earth-based stations and  therefore the nominal diffraction-limited angular resolution for a given wavelength $\lambda$ is limited by the Earth's diameter $D$. For observations at 230 GHz, this is approximately $\lambda/D\approx 21 \, \mu as$ and the maximum possible baseline length in units of wavelength, is $D/\lambda\approx 9.6 $\,G$\lambda$. In order to improve the limitations on dynamical range and temporal resolution arising due to EHT's sparse $(u,v)$ coverage \citep{EHT_M87_II}, as well to increase the achievable baseline lengths, there are several next-generation black hole imaging  missions potentially offering sensitivity to the photon ring signals. The next-generation Event Horizon Telescope (ngEHT) is a ground-based extension to the EHT that aims to install new telescope sites, expand the observing bandwidth to 16 GHz and perform simultaneous multi-frequency observations at 86-230-345 GHz, all of which can lead to high-fidelity, real-time movies of black holes and their associated jets \citep{Doeleman:2023}. The Black Hole Explorer (BHEX) mission \citep{Johnson:2024}, aims to perform Earth-space VLBI with a single space-based orbiter, providing long baselines $(\gtrapprox 20 G\lambda)$ enabling photon ring science while also obtaining a dense coverage of the (u,v) plane on short baselines as well, with the aid of simultaneous dual-band observations by leveraging the expansions on the ground.

To study the efficacy of these missions to probe the photon ring, this paper studies its polarimetric features through both, geometric modelling and time-averaged GRMHD simulations. Indeed, while geometric modelling allows to probe lensing structures that are largely agnostic to astrophysical source profiles \citep{Gralla:2020_grt}, simulations help identify features of the accretion flow that are strongly sensitive to the presence of the photon ring \citep{Ricarte:2021,Rosales:2021}. The geometric modelling of photon ring signatures of \m{} has been studied by several authors in total intensity \citep{Johnson_2020,Gralla:2020_grt,Lockhart:2022A,Paugnat:2022,Avendano:2023,Jia:2024}, and in LP \citep{Himwich:2020}. \cite{Gralla:2020_shape} has also studied signatures on long baselines of general narrow features of arbitrary shape and intensity. The polarimetric signature of \m{}'s photon ring in GRMHD simulations has also been studied by several authors \citep{Ricarte:2021,MM_2021,Rosales:2021,Palumbo:2023}.


The paper is organised as follows. Section \ref{sec:interf_rings} introduces the geometric model of the ring. Section \ref{sec:ring_with_rho_formulas} explores the photon ring transition signatures of the geometric model in all polarizations. Section \ref{sec:GRMHD} analyses transition behavior in a large library of GRMHD simulations in light of the geometric analysis. Section \ref{sec:instrumentation} describes detection requirements for ngEHT and BHEX in terms of permissible thermal noise and antenna diameter. Section \ref{sec:conclusions} presents the conclusions and discusses the outlook for future work. Extended derivations of formulae used in the paper, as well as tabulated numerical results from GRMHD, can be found in the appendices.


\section{Interferometric Signatures of Analytic Ring Models} \label{sec:interf_rings}

\begin{figure*}[t!]
    \centering
    \includegraphics[width=\textwidth]{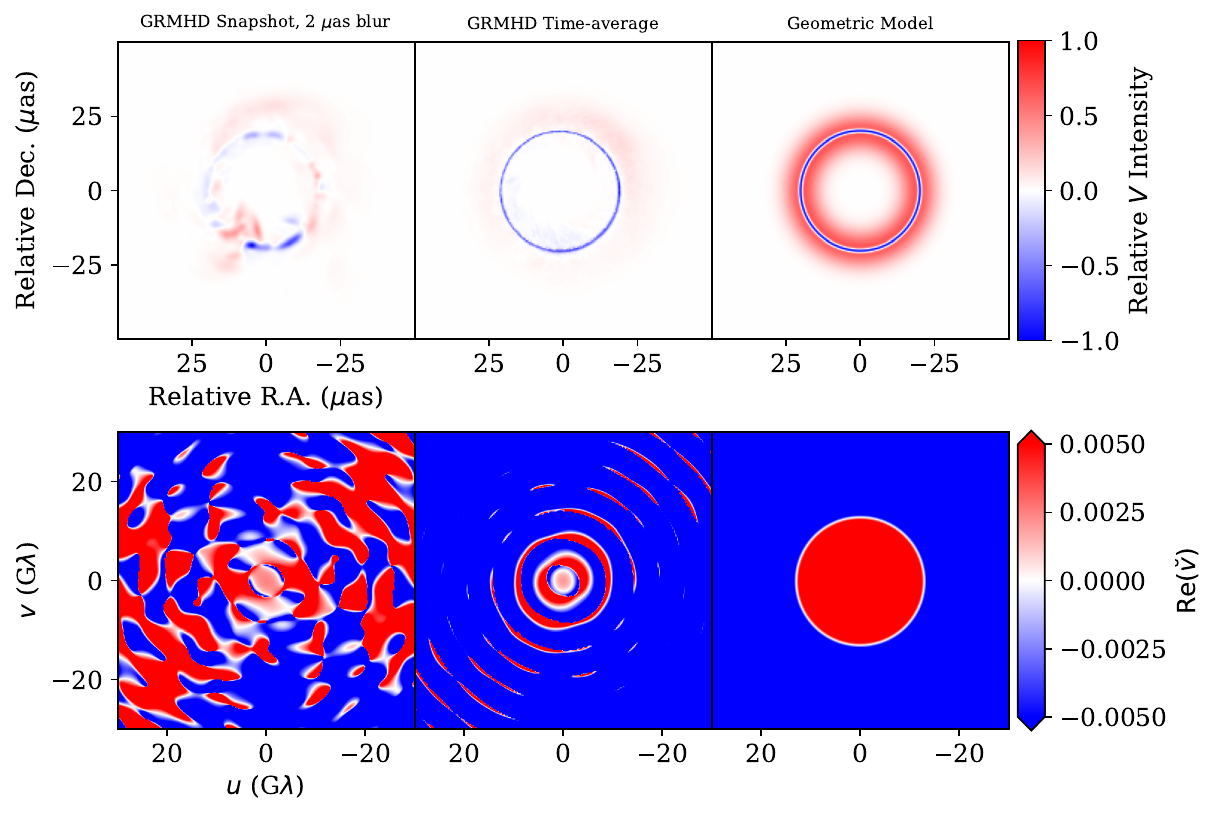}
    \caption{Example of Stokes $V$ transition in GRMHD as motivation for the geometric ring model. Top left: Stokes $V$ snapshot from a MAD GRMHD simulation with $a_*=-0.5$, a viewing inclination of $17^\circ$, and electron heating parameter $R_{\rm h}=160$. Top middle: time-averaged Stokes $V$ image from the same model. Top right: geometric model of two concentric rings of equal radius but different Gaussian thicknesses. Bottom row: corresponding real part of the interferometric quotient $\breve{v} \equiv \tilde{V} / \tilde{I}$. The transition from direct-image dominated (red) to photon ring-dominated (blue) occurs at a radius in the $(u,v)$ plane dependent on model parameters such as ring thickness and relative brightness; in this example, the GRMHD model's photon ring causes a transition to negative $\breve{v}$ on shorter baselines than this example geometric model.}
    \label{fig:modelintro}
\end{figure*}

Let the image plane be spanned by some two dimensional co-ordinate system $\vec{r}=(x,y)$ and the baseline between the two stations spanned by a dimensionless vector $\vec{b}=(b_{1},b_{2})$ drawn perpendicular to the line of sight. Then, the visibility response $V(\vec{b})$ of an interferometer formed by this baseline, for an image signature $I(\vec{r})$, is given by a two-dimensional Fourier transform:
\begin{gather}
    \tilde{V}(\vec{b})=\int I(\vec{r})e^{-2\pi i \vec{b}.\vec{r}}d^{2}\vec{r}. \label{eq:visibility}
\end{gather}
Here the ``tilde" ($\tilde{V}$) denotes a quantity in the Fourier/visibility domain. Note that this relation applies for all Stokes parameters $(I,Q,U,V)$ where $I$, $P=Q+iU$ and $V$ represent total intensity, LP and CP respectively. 

We consider the visibility domain to be spanned by polar co-ordinates $(\rho,\theta)$. These co-ordinates are related to the Cartesian $(u,v)$ co-ordinates via the relation:
\begin{gather}
    \rho=\sqrt{u^{2}+v^{2}},\quad \theta=\tan^{-1}\bigg(\frac{u}{v}\bigg).
\end{gather}

\subsection{The Model: Concentric, Gaussian Rings}

We now specify a geometric model of the image in terms of concentric, Gaussian rings. Such constructions have been used by the EHT to model-fit observations and study the effects on ring parameters of finite imaging resolution \citep{EHT_M87_IV}. The image model is based on the following assumptions:
\begin{enumerate}
    \item The model for Stokes $I$ begins with a pair of symmetric, infinitesimally thin rings, each following \citet{Johnson_2020} for a given sub-image $n$:
    \begin{align}
        I_n(r)&= \frac{1}{2 \pi r_n} \delta\left(r-r_n\right).
    \end{align}
    \item We assume hierarchical total fluxes and thicknesses for the two rings in the model. The thicker ring, corresponding to the $n=0$ signature, has a total flux $F_{0}$, radius $r_{0}$ and a thickness modelled by a Gaussian of thickness (standard deviation) $\sigma_{0}$. Similarly, the thinner ring corresponding to the $n=1$ signature has a total flux $F_{1}$, radius $r_{1}$ and Gaussian thickness $\sigma_{1}$. Recall that the Gaussian thickness $\sigma$ is related to the Full Width at Half Maximum (FWHM) by the relation, 
    \begin{gather}
        \theta=2\sqrt{2\text{ln}2}\sigma \label{eq:fwhm_sigma}.
    \end{gather}
    \item The signature for $n=0$ and $n=1$ in Stokes V is proportional to that of Stokes I, with uniform fractional circular polarization $v_0$ and $v_1$ for the $n=0$ and $n=1$ images, respectively. That is, for a given sub-image $n$,
    \begin{align}
        V_n(\rho,\theta) &= v_n I_n(\rho, \theta).
    \end{align}
    \item The LP signature is also proprtional to that of Stokes I scaled version of the Stokes I signature in the image domain \citep{Johnson_2020,Palumbo:2020} with scaling described by factors $\beta_{2,0}$ and $\beta_{2,1}$ for the $n=0$ and $n=1$ rings respectively, along with the exponential $e^{i2\theta}$ to produce rotationally symmetric electric vector position angle. That is, for a given sub-image $n$,
    \begin{align}
        P_n(\rho, \theta) &= \beta_{2,n} e^{i2\theta}I_n(\rho, \theta).  
    \end{align}
\end{enumerate}
The total signature in the Fourier domain is a sum of the $n=0$ and $n=1$ visibility responses:
\begin{align}
    \tilde{I}_{01}&=F_{0}J_{0}(2\pi r_{0}\rho)e^{-2\pi^{2}\sigma_{0}^{2}\rho^{2}}\nonumber \\
    &+F_{1}J_{0}(2\pi r_{1}\rho)e^{-2\pi^{2}\sigma_{1}^{2}\rho^{2}}, \label{eq:I_01}\\
    \tilde{P}_{01}&=-\beta_{2,0}F_{0}e^{2i\theta}J_{2}(2\pi r_{0}\rho)e^{-2\pi^{2}\sigma_{0}^{2}\rho^{2}}\nonumber \\ &-\beta_{2,1}F_{1}e^{2i\theta}J_{2}(2\pi r_{1}\rho)e^{-2\pi^{2}\sigma_{1}^{2}\rho^{2}},\label{eq:P01}\\
    \tilde{V}_{01}&=v_{0}F_{0}J_{0}(2\pi r_{0}\rho)e^{-2\pi^{2}\sigma_{0}^{2}\rho^{2}}\nonumber \\ &+v_{1}F_{1}J_{0}(2\pi r_{1}\rho)e^{-2\pi^{2}\sigma_{1}^{2}\rho^{2}}. \label{eq:V_01}
\end{align}
Here $J_{m}$ is Bessel function of the first kind of order $m$. For the subsequent analysis, we introduce the notation for the image domain ratios flux and polarimetric ratios,
\begin{align}
\mathcal{F}=\frac{F_{0}}{F_{1}},\quad \beta_{r}=\frac{\beta_{2,0}}{\beta_{2,1}},\quad v_{r}=\frac{v_{0}}{v_{1}}, \label{eq:notation}
\end{align}
the Fourier domain polarimetric ratios,
\begin{align}
\breve{m}_{01}=\frac{\tilde{P}_{01}}{\tilde{I}_{01}},\quad\breve{v}_{01}=\frac{\tilde{V}_{01}}{\tilde{I}_{01}},
\end{align}
and the LP quantity,
\begin{align}\bar{m}_{01}=\breve{m}_{01}(e^{-2i\theta}). \label{eq:lp_quantity}
\end{align}
Further details of the model can be found in Appendix \ref{sec:model_theory}. Figure \ref{fig:modelintro} demonstrates how this model approximates the polarized photon ring features of the image domain and the corresponding interferometric transition structures that manifest more subtly in GRMHD simulations.

We further restrict the model by fixing several parameter ranges relative to each other to simplify the model specification and mimic  properties of the photon ring often seen in simulations of the \m{} and \s{} accretion flows:
\begin{enumerate}
    \item We fix $F_{0}=1$ and consider the flux ratio $\mathcal{F}$ to be a free parameter.
    \item Based on insights from \citet{Palumbo:2022}, the $\beta_{2}$ quantities are assumed to satisfy the complex conjugacy relation
    \begin{gather}
        \beta_{2,1}=\beta_{2,0}^{\star}\Rightarrow \text{arg}(\beta_{2,1})=-\text{arg}(\beta_{2,0})\label{eq:b2_relation},
    \end{gather}
    and so $\beta_{2,0}$ and the ratio $\beta_{r}$ are free parameters.
    \item We assume the fractional circular polarizations $v_0$ and $v_1$ have opposite sign, and take $v_{0}$ and the ratio $v_{r}$ to be free parameters.
    
    \item We assume that the two rings have equal radius. This is physically motivated from studies of thin ring models in GRMHD simulations \citep{Tiede:2022} as well as from the analytical work by \cite{Ozel:2022} which arrived at similar results using a covariant model of the accretion flow.
\end{enumerate}
Thus the specifiable parameters of the model (excluding the ring radii which are discussed in the next section) are: $\mathcal{F}, \beta_{2,0},\beta_{r},\sigma_{0},\sigma_{0}/\sigma_{1},v_{0},$ and $v_{r}$.

\section{Polarimetric Photon Ring Signatures: Concentric Gaussian Rings} \label{sec:ring_with_rho_formulas}
We now study the transition in $\rho$ at which the $n=1$ signature overtakes the $n=0$ signature in each polarization. Though we consider rings of equal radii for the sake of completeness, an analytic description of such transitions without using this assumption is provided in the Appendix \ref{sec:transition_deriv}. The calculation for CP is given below, while for total intensity and LP is given in Appendix \ref{sec:transition_deriv}.

For CP, using Equations \ref{eq:I_01} and \ref{eq:V_01}, we get:
\begin{gather}
\breve{v}_{01}=\frac{v_{0}F_{0}J_{0}(2\pi r_{0}\rho)e^{-2\pi^{2}\sigma_{0}^{2}\rho^{2}}+v_{1}F_{1}J_{0}(2\pi r_{1}\rho)e^{-2\pi^{2}\sigma_{1}^{2}\rho^{2}}}{F_{0}J_{0}(2\pi r_{0}\rho)e^{-2\pi^{2}\sigma_{0}^{2}\rho^{2}}+F_{1}J_{0}(2\pi r_{1}\rho)e^{-2\pi^{2}\sigma_{1}^{2}\rho^{2}}}\label{eq:vbreve_general}.
\end{gather}
Imposing the condition of equal radii cancels out $J_{0}(...)$ terms, and taking $v_{0}F_{0}e^{-2\pi^{2}\sigma_{0}^{2}\rho^{2}}$ and $F_{0}e^{-2\pi^{2}\sigma_{0}^{2}\rho^{2}}$ common from the numerator and denominator respectively gives:
\begin{gather}
    \breve{v}_{01}=v_{0}\Bigg(\frac{1+\frac{v_{1}F_{1}}{v_{0}F_{0}}e^{2\pi^{2}(\sigma_{0}^{2}-\sigma_{1}^2)\rho^{2}}}{1+\frac{F_{1}}{F_{0}}e^{2\pi^{2}(\sigma_{0}^{2}-\sigma_{1}^2)\rho^{2}}}\Bigg).
\end{gather}
The condition for $\breve{v}_{01}$ to have a negative sign is,
\begin{gather}
    \bigg\lvert \frac{v_{1}}{v_{0}}\bigg\rvert \bigg(\frac{F_{1}}{F_{0}}\bigg)e^{2\pi^{2}(\sigma_{0}^{2}-\sigma_{1}^{2})\rho^{2}}>1    \label{eq:sign_vbrv_cond}.
\end{gather}
Solving Equation \ref{eq:sign_vbrv_cond} for $\rho$ and using the notation from Equation \ref{eq:notation} gives the transition value $\rho\rightarrow(\rho_{T})_{V}$ at which the sign changes as,
\begin{gather}
    (\rho_{T})_{V}=\sqrt{\frac{\ln\big(\lvert v_{r}\rvert\mathcal{F}\big)}{2\pi^{2}(\sigma_{0}^{2}-\sigma_{1}^{2})}}. \label{eq:cp_trans}
\end{gather}

A key insight that arises from Equation \ref{eq:sign_vbrv_cond} is that the physical condition of the $n=1$ signal dominating over $n=0$, is \textit{exactly} equivalent to the mathematical condition for $\breve{v}_{01}$ to change its sign. This further motivates the subsequent analysis of the GRMHD simulations to probe the photon ring signatures in these quantities.

Moreover, noting the fact that in Equation \ref{eq:cp_trans} the quantities inside the square root are all ratios and hence dimensionless, it can be inferred that the strongest determinant of $\rho_T$ is the ``resolving out'' of the direct image by exceeding $\rho = 1/\sigma_0$. This is evident from rewriting Equation \ref{eq:cp_trans} as:
\begin{gather}
    (\rho_{T})_{V}=\frac{1}{\sigma_{0}}\sqrt{\frac{\ln\big(\lvert v_{r}\rvert\mathcal{F}\big)}{2\pi^{2}\Big(1-\frac{\sigma_{1}^{2}}{\sigma_{0}^{2}}\Big)}}. \label{eq:sigma_vbrv}
\end{gather}

The thickness of the ring is is a crucial factor in comparing simulations with observations \citep{Lockhart:2022,Lockhart:2022A} and so Equation \ref{eq:sigma_vbrv} and can be a useful tool for geometric modelling of polarimetric signatures of the photon ring. Moreover, the ensuing geometric features can be explored solely by fixing the \textit{one} parameter of the $n=0$ ring, namely $\sigma_{0}$, and then defining the \textit{ratios} of quantities relating the $n=0$ and $n=1$ ring.

The formulae for total intensity and LP, reproduced from Equations \ref{eq:rho_t_I} and \ref{eq:rho_t_mbreve} with the notation of Equation \ref{eq:notation}, are given by:
\begin{gather}
    (\rho_{T})_{I}=\sqrt{\frac{\ln(\mathcal{F})}{2\pi^{2}(\sigma_{0}^{2}-\sigma_{1}^{2})}}, \label{eq:i_trans}\\(\rho_{T})_{\text{LP}}= \sqrt{\frac{\text{ln}(|\beta_{r}|\mathcal{F})}{2\pi^{2}(\sigma_{0}^{2}-\sigma_{1}^{2})}} \label{eq:lp_trans}.
\end{gather}

\begin{figure*}
    \centering    \includegraphics[width=0.99\textwidth]{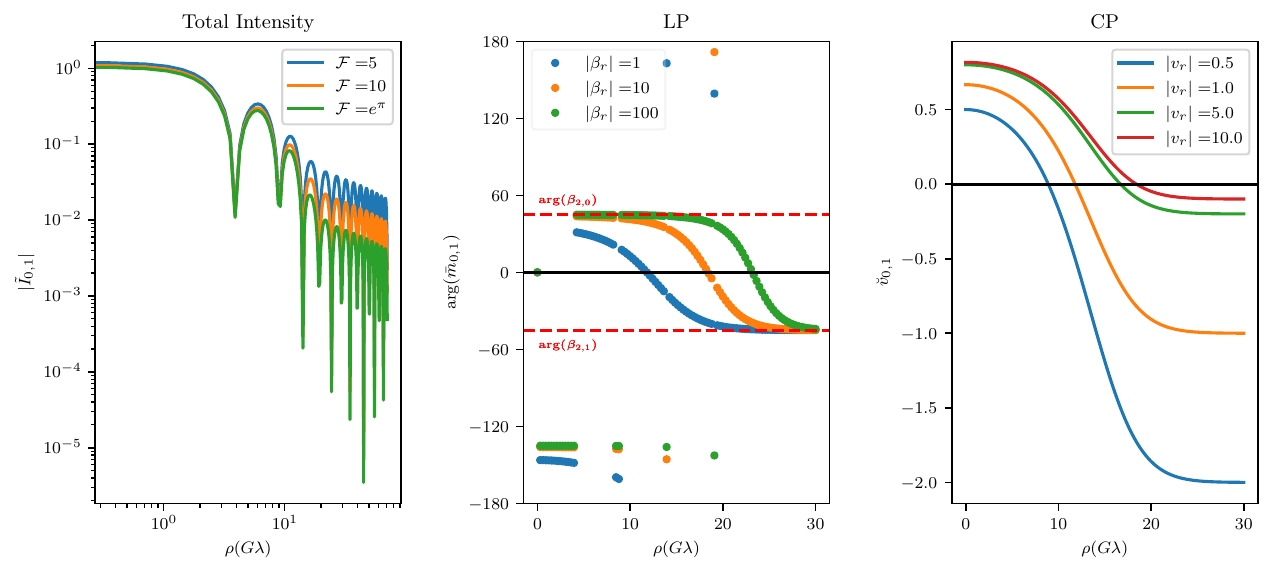}
    \caption{The behaviour of total intensity (on log scale), LP and CP as a function of baseline length $\rho$, measured in units of Giga $\lambda$ (G$\lambda$). Here $\bar{m}_{01}=\breve{m_{01}}e^{-2i\theta}$ and arg($\beta_{2,1})=-$arg($\beta_{2,0})=45^{\circ}$, with the latter condition arising from Equation \ref{eq:b2_relation}. For all the plots, it is assumed that $\sigma_{0}=5\mu$as and $\sigma_{0}/\sigma_{1}=15$. For the LP and CP plots, it is assumed that $\mathcal{F}=5$.}
        \label{f:ipv_plot}
\end{figure*}

Using Equations \ref{eq:cp_trans}, \ref{eq:i_trans} and \ref{eq:lp_trans}, Figure \ref{f:ipv_plot} shows example interferometric quantities as a function of $\rho$ for combinations of model parameters that vary the relative contribution of the photon ring. We investigate the visibility domain signatures in $\tilde{I}_{01}, \text{arg}(\breve{m}_{01}e^{-2i\theta})\equiv\text{arg}(\bar{m}_{01})$ and $\breve{v}_{01}$. We include the ratio $\mathcal{F}\approx e^{\pi}$ to mimic the ``universal regime'' prediction for isotropic emitters \citep{Johnson_2020} and to make contact with the results of \cite{Tiede:2022} for \m. The Figure demonstrates the advantages of the geometric modelling scheme since after fixing geometric parameters of the rings (i.e $\sigma_{0}$ and $\sigma_{0}/\sigma_{1}$),  the onset of the photon ring transition is primarily governed by the relative polarization ratios ($\mathcal{F}$, $|\beta_{r}|$ and $|v_{r}|$) of the $n=0$ to the $n=1$ image. We find  that the the photon ring signal starts to dominate at longer baselines ($\gtrapprox 10 G\lambda)$, tuned by model parameters. Specifically for 
LP, as the $n=1$ ring becomes more linearly depolarised with respect to the $n=0$ ring (i.e as $|\beta_{r}|$ increases), the transition radius $\rho_{T}$ at which the photon ring begins to dominate occurs at longer baselines. For CP, similar trends are observed for the values of the ratio $|v_{r}|$.

\section{Comparison with GRMHD Simulations}\label{sec:GRMHD}
We now investigate the range of values of $\rho_{T}$ for GRMHD simulations of \m, focusing on LP and CP signatures. We consider a library of time-averaged ray-traced GRMHD $n=0$ and $n=1$ images produced in \citet{Palumbo:2022}, which hold fixed the mass at $6.5\times10^{9} M_{\odot}$, the distance at $16.8$ Mpc, and the inclination at $17^\circ$. The library contains five values of dimensionless spin $a_*$: -0.94, -0.5, 0, +0.5, and +0.94, where a minus/plus sign indicates retrograde/prograde spin with respect to the large-scale accretion disk. In all cases, the black hole spin is directed away from the observer. The fluid models themselves were produced by \texttt{iharm3D} \citep{Gammie_HARM_2003, IHARM3d_prather} and ray-traced using \texttt{IPOLE} \citep{IPOLE_2018} with six values of the electron heating parameter $R_{\rm h}$ \citep{Mosci_2016}: 1, 10, 20, 40, 80, and 160. This parameter tunes the relative temperatures of ions and electrons as a function of the magnetization of the plasma.


\begin{figure*}[t!]
    \centering
    \includegraphics[width=\textwidth]{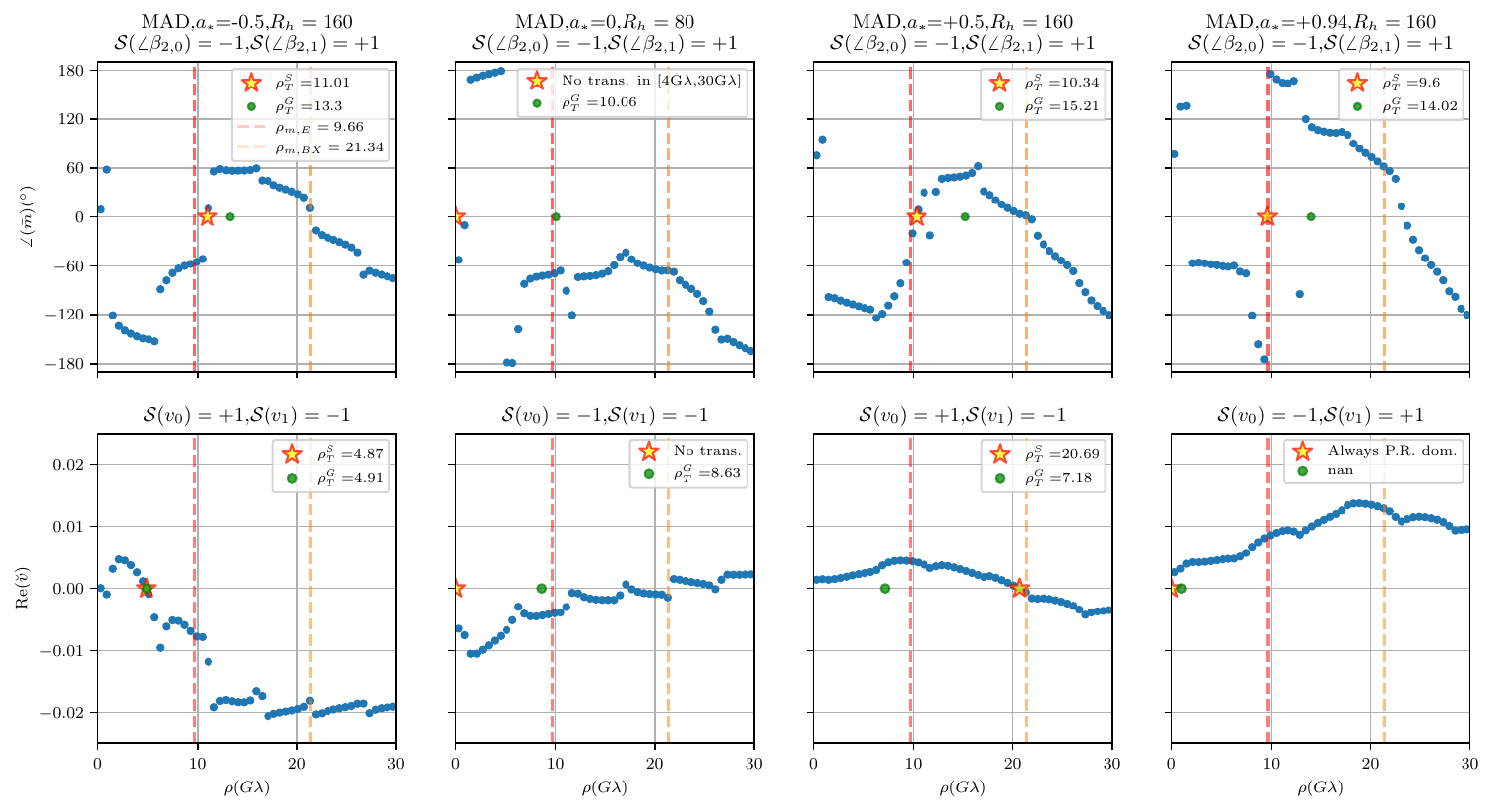}
    \caption{The signatures in $\angle(\bar{m})$ and Re$(\breve{v})$ for the four polarimetric best-bet models at 230 GHz for M87*. The ``stars" represent the transition radii obtained from the change in sign (denoted by $\mathcal{S}$) from the $n=0$ to the $n=1$ image domain polarization quantity. The ``dots" represent the values obtained from the geometric formulae developed in the paper. For MAD, $a_*=0$, $R_{\rm h}=80$, no transition is reported in LP since the physical transitions are posited to occur beyond 30 G$\lambda$. For MAD, $a_*=+0.94$, $R_{\rm h}=160$ since $\mathcal{S}(v_{1})=1$ and the signal only has positive values, the inference for the CAA-based analysis is that the signal is always photon ring dominated. The nan represents the mathematical fact that for this model, $\mathcal{F}\times v_{r}<1$ and so Equation \ref{eq:cp_trans} has an undefined numerical value.}
        \label{f:best_bet}
\end{figure*}
\subsection{Methodology: Coherent Annular Averaging}
\begin{figure*}
    \centering    \includegraphics[width=0.99\textwidth]{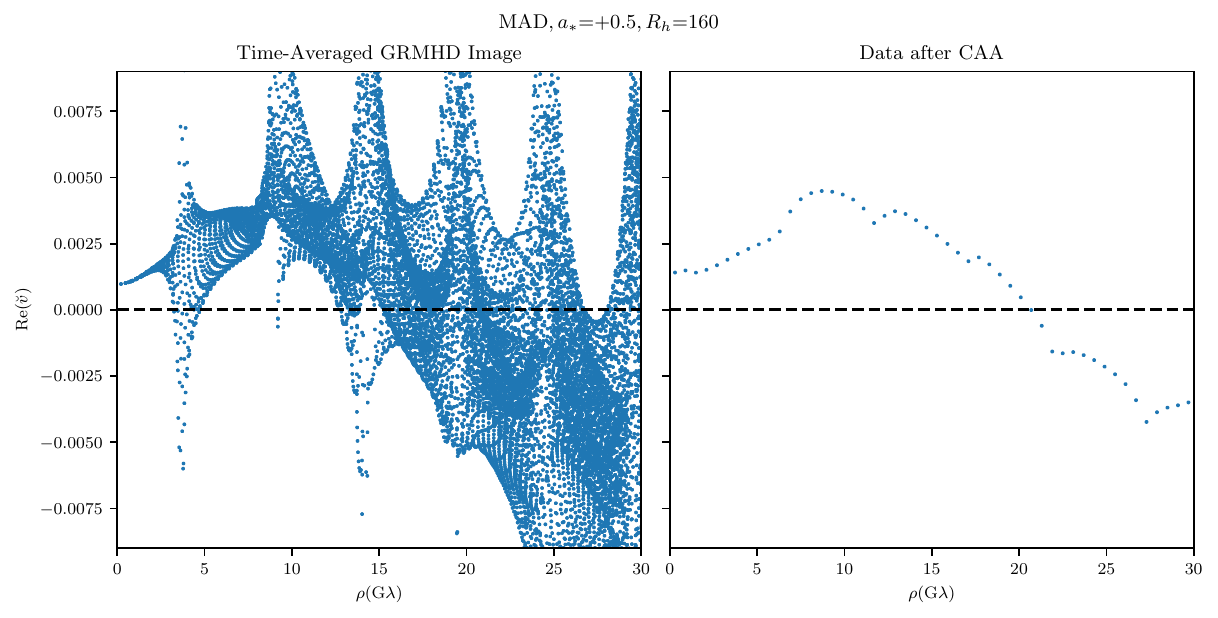}
    \caption{The demonstration of the CAA scheme on a time-averaged GRMHD image of a MAD model with $a_{*}=+0.5$ and $R_{h}=160$ which contains contributions from both $n=0$ and $n=1$ images. The left panel shows Re($\breve{v}$) for a grid of points in the $(u,v)$ plane out to 30 G$\lambda$ while the right panel shows the coherently averaged results after implementing CAA with a bin width of 4$\,G\lambda$. }
        \label{f:caa_plot}
\end{figure*}
In order to obtain the transition radii directly from the time-averaged images, 
we utilise a coherent annular averaging (CAA) scheme. Herein, for each image, a  polarimetric quantity  is studied by making annular averages using a bin width of the order of few G$\lambda$. Throughout this analysis, a bin width of 4G$\lambda$ is chosen. Physically, this corresponds to smoothing the small-scale fluctuations in the Fourier domain signal to investigate at which transition radii the photon ring signal \textit{first} begins to dominate over the direct image. To specify this in the signal, we utilise the change in sign from the $n=0$ to the $n=1$ image-domain polarimetric quantities. For LP and CP, we track the sign change from $\beta_{2,0}$ to $\beta_{2,1}$ and $v_{0}$ to $v_{1}$ respectively. In practice, the CAA-based inference scheme is implemented as follows: 
\begin{enumerate}
    \item For a fixed value of the bin width $b$, specify the inner ($\rho_{-}$) and outer ($\rho_{+}$) radii of the annulus in the Fourier domain:
    \begin{gather}
        \rho_{\pm}=\rho\pm\frac{b}{2}.
    \end{gather}
    Then, for each value of $\rho$ in an equally-spaced array from 0 to (say) $30 $\,G$\lambda$, extract the points $\rho_{\text{annulus}}$ in the corresponding annuli by performing a boolean \& operation as, 
    \begin{gather}
        \rho_{\text{annulus}}=(\rho<\rho_{+})\&(\rho>\rho_{-}).
    \end{gather}
    \item Mask the interferometric quantity using the $\rho_{\text{annulus}}$ values.
    \item Take the mean of the resultant masked values and plot them as a function of $\rho$. These plots are then used to investigate aforementioned transition beyond which the photon ring begins to dominate.
\end{enumerate}

A demonstration of CAA on a time-averaged GRMHD image is given in Figure \ref{f:caa_plot}, wherein a MAD model with $a_{*}=+0.5$ and $R_{h}=160$ is studied. The left panel demonstrates Re($\breve{v}$) plotted as a function of $\rho$ when no averaging scheme has been applied. The right panel demonstrates the data obtained after implementing CAA. The latter shows a distinct signature in the Fourier plane of a transition point beyond which the character of the signal changes. In subsequent sections, this will be identified as demonstrating the onset of the regime where the photon ring signal begins to dominate over the direct image.

The implementation of CAA is particularly important for $\breve{m}$ due to the phase swings introduced by the $J_2-J_0$ null separation \citep{Palumbo:2023}, which are smoothed over by this averaging. Similarly for $\breve{v}$, CAA averages out the spurious phase wraps that arise from the Stokes V and I images not necessarily being exactly proportional to one another. 

\subsection{Results}
To demonstrate the signal characteristics in LP and CP using CAA, we first consider the four models for M87* at 230 GHz identified by the recent CP results of the EHT \citep{EHT_M87_IX} as having the strongest agreement with EHT polarimetric data constraints; note that these models were not compared to other external criteria, such as jet power. These are MAD models with spin and $R_{\text{h}}$ values of $(a_*=-0.5,R_{\rm h}=160)$, $(a_*=0,R_{\rm h}=80)$, $(a_*=+0.5,R_{\rm h}=160)$ and $(a_*=+0.94,R_{\rm h}=160)$. The transition radii from the CAA scheme as well as from the geometric formulae are given in Figure \ref{f:best_bet}. In the LP plots, the baselines shorter than 4 G$\lambda$ are avoided based on insights from \citet{Palumbo:2023} to prevent picking up the relative sign difference between the Bessel functions $J_{2}$ and $J_{0}$ while the upper limit of 21 G$\lambda$ is chosen since this is approximately the maximum baseline of BHEX at 230 GHz. For CP, since there are no such sign complications, the lower and upper bounds are chosen to be 1 G$\lambda$ and 21 G$\lambda$ respectively. The CP signal for MAD model with $a_{*}=0$ and $R_{h}=160$ is peculiar since here Stokes V signal has the same sign of $v_{0}$ and $v_{1}$ which is inconsistent with our criteria our assigning a change in sign to indicate dominance of the photon ring over the direct image.  

\begin{figure*}[t!]
    \centering
    \includegraphics[width=0.99\textwidth]{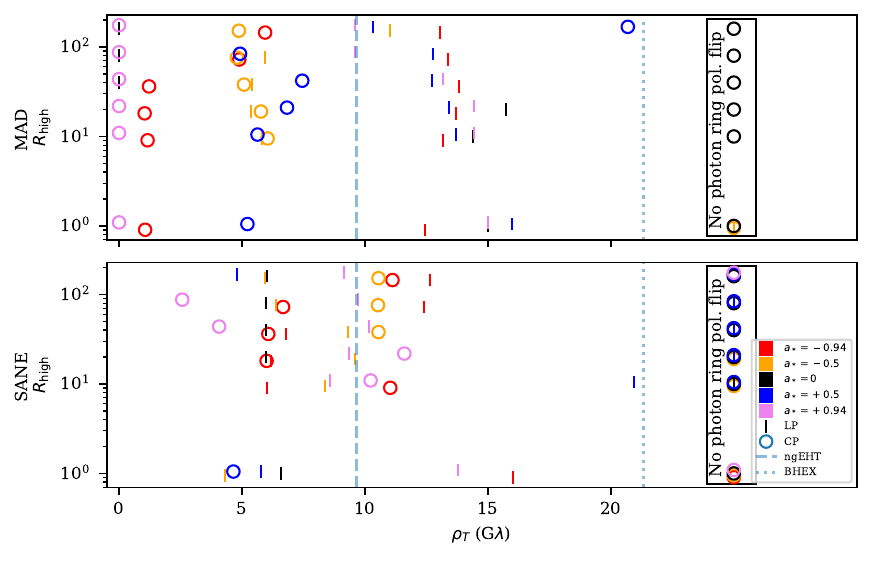}
    \caption{The transition values $\rho_{T}$ in the Fourier plane obtained via CAA of time-averaged GRMHD images of M87* at 230 GHz. The data points denote the \textit{first} instance at which the photon ring signal begins to dominate. For LP, this is noted by identifying the value of $\rho$ at which the sign of arg($\bar{m}_{01}$) swings from positive to negative or vice versa, with the ``correct" swing being decided by the whether the corresponding change in signs of arg$(\beta_{2,0})$ to arg$(\beta_{2,1})$ (which are image domain quantities) is positive to negative or vice versa. For CP, a similar prescription is provided for noting the sign change in the real part of $\breve{v}_{01}$, with the ``correct" convention based on the signs of $v_{0}$ and $v_{1}$. Markers in the black rectangle on the right indicate images for which the direct and indirect images have the same polarimetric signs and therefore do not show a clear photon ring transition in the indicated polarized interferometric quantity.}
        \label{f:grmhd_summary}
\end{figure*}
Next, we apply the CAA scheme to thehe full suite of GRMHD simulatio. The a graphic summary of the transition radii obtained is provided in Figure \ref{f:grmhd_summary} and the exact values are tabulated in Table \ref{t:grmhd_geometric_vals}, which also contains the transition radii for LP and CP for each model obtained from the geometric formulae given in Equations \ref{eq:lp_trans} and \ref{eq:cp_trans} respectively. To obtain the values from the geometric formulae, the ring widths were obtained using \textbf{\texttt{VIDA.jl}} \citep{Tiede:2020} by fitting a second-order $m$-ring model onto the image.

\subsubsection{Inferences and Trends}
From the values obtained in Table \ref{t:grmhd_geometric_vals}, several consistency checks can be performed with previous polarimetric studies of the photon ring. For example, for MAD states with a given spin, as $R_{h}$ increases, $|\beta_{r}|$ increases, implying a strong depolarization of the $n=1$ photon ring relative to the $n=0$ ring. This is consistent with the findings of \citep{Palumbo:2022}, which used the same dataset as this paper. Furthermore, \cite{Palumbo:2023} studied the MAD model with $a_*=+0.5$, $R_{h}=80$ and found a clear photon ring transition in LP at around $15$G$\lambda$, which is reasonably close to the values obtained from the LP geometric model ($14.02$ G$\lambda$) and CAA ($12.75$ G$\lambda$). 

However, trends with respect to $R_{\rm h}$ are not in general simple. Though increasing $R_{\rm h}$ tends to produce more jet emission in both MADs and SANEs, the degree to which it also increases the population of cold, depolarizing electrons is generally higher in MADs \citep{EHT_M87_VIII}. Abstractly, changing $R_h$ changes the relative balance of different polarizing and depolarizing mechanisms by changing the emission geometry and electron temperature, potentially leaving the simple regime of uniform structures in $n=0$ simply negated in the $n=1$ image. Though other electron thermodynamical post-processing schemes could produce systematically different transition regions this way, the existence of a transition in the Fourier domain can only be undone by a drastic change in optical or Faraday depths, which is difficult to imagine without breaking data consistency with EHT observations.

We now focus on the tabulated values obtained by applying CAA to the GRMHD simulations suite. For the favoured MAD state of M87* \citep{EHT_M87_VIII}, the values given in column $\rho_{L}^{S}$ and $\rho_{C}^{S}$ allows us to observe several spin-sensitive trends for the LP and CP transition radii respectively. In LP, for retrograde spins, increasing the spin value from $a_*=-0.5$ to -0.94 while keeping $R_{\text{h}}$ fixed pushes all the photon ring transitions beyond maximum Earth baselines at 230 GHz. The two outliers here are models with spin $a_*=-0.5$, with $R_{\text{h}}=1$ not having a transition and $R_{\text{h}}=160$ being already beyond the maximal Earth baseline. Next, for $a_*=|0.5|$, the transition radii values are quite sensitive to whether the spin is prograde/retrograde. All negative spin values are below the maximum Earth baselines (except $R_{\text{h}}=160$) while positive spin values are beyond Earth baselines. Once again $a_*=-0.5$ and $R_{\text{h}}=1$ is an outlier. The models with spins $a_*=|0.94|$ \textit{do not} have this sensitivity to the direction of the spin. However, while all $a_*=|0.94|$ models have transition radii in LP that are beyond the maximal Earth baselines at 230 GHz, the model with $a_*=+0.94$ and $R_{\text{h}}=160$ has a transition radius that is \textit{just on the edge} of this maximal length. We have checked that this result holds even when the bin size is $2 $\,G$\lambda$ and so there is some confidence that the transition value is not an artefact of ``over-smoothing" of the data. 

For the values in CP, most MAD models have transition radii that are accessible from Earth baselines. Furthermore,  MAD models with $a_*=+0.94$ and $R_{\rm h}>1$ all produce the majority of their CP flux in the photon ring with a sign flip, causing the photon ring to dominate on all baselines. The mathematical equivalent of this result from the geometric formula is that with the exception of the model with $R_{\rm h}=1$, all MAD models with $a_*=+0.94$ have $\mathcal{F}\times v_{r}<1$ and so the logarithm inside the square root in Equation \ref{eq:cp_trans} is undefined. Besides this spin, the value $\rho_{C}^{S}=0.0$ in all other cases can be explained by the absence of a sign change $\mathcal{S}(v_{0})\rightarrow\mathcal{S}(v_{1})$ thereby implying the non-existence of a transition from the $n=0$ to the $n=1$ regime.

Lastly, we note the particularly interesting case of MADs with $a_*=0$. Here, for all values of $R_{h}$, there is no transition in the signs of $v_{0}$ and $v_{1}$ (i.e both $\mathcal{S}(v_{0})=\mathcal{S}(v_{1})=-1$) and so no transition radii in CP are obtained from the simulations. The absence of such a trend in any other spin values for MAD models indicates the potential utility of CP photon ring signatures to assert a non-zero spin value for \m{}. This can complement the recent LP-based analysis by \cite{Chael:2023} of M87*'s jet using the Blandford-Znajek mechanism, which crucially requires the black hole to have a non-zero spin.


\section{Instrumentation Considerations for Polarimetric Best-Bet Models} \label{sec:instrumentation}
In order to address whether ngEHT and BHEX can indeed observe the LP and CP photon ring signatures discussed above, we focus on the four polarimetric best-bet models and consider two main quantities of interest, namely the maximum thermal noise and the antenna size. The thermal noise is a proxy for sensitivity requirements of a baseline while greater antenna diameter implies better performance of the given station.

\subsection{Maximum Permissible Thermal Noise}
To introduce the notion of maximum permissible thermal noise, we first consider the case for CP. Under the assumption that the thermal noise for $\tilde{V}$ and $\tilde{I}$ is equal, the thermal noise $\sigma_{\breve{v}}$ for their quotient $\breve{v}$ is given by,
\begin{align}
    \frac{\sigma_{\breve{v}}}{|\breve{v}|}=\sigma\sqrt{\frac{1}{|\tilde{V}|^{2}}+\frac{1}{|\tilde{I}|^{2}}}.
\end{align}
Since we require the left hand side to be less than 1, the maximum value $\sigma_{\text{max},\breve{v}}$ would be obtained by equating the right hand side to 1 and solving for $\sigma\rightarrow\sigma_{\text{max},\breve{v}}$. This gives,
\begin{gather}
    \sigma_{\text{max},\breve{v}}=\frac{1}{\sqrt{\frac{1}{|\tilde{V}|^{2}}+\frac{1}{|\tilde{I}|^{2}}}}. \label{eq:cp_max_tn}
\end{gather}
For LP, since $\theta$ in Equation \ref{eq:lp_quantity} is a function of $u$ and $v$ co-ordinates which can be defined very accurately, the error in it will be only due to the $\breve{m}_{01}$ term. Once again, assuming that thermal noise in $\tilde{Q}$ and $\tilde{U}$ is equal, in the high signal-to-noise limit \citep{Chael:2016}, the expression for the maximum thermal noise can be obtained analogous to the case for CP, and is given by
\begin{align}
    \sigma_{\text{max},\breve{m}}=\frac{1}{\sqrt{\frac{2}{|\tilde{P}|^{2}}+\frac{1}{|\tilde{I}|^{2}}}} \label{eq:lp_max_tn}.
\end{align}
An extended derivation of Equations \ref{eq:cp_max_tn} and \ref{eq:lp_max_tn} is given in Appendix \ref{sec:sensitivity_tn}.

\begin{figure*}
    \centering
    \includegraphics[width=\textwidth]{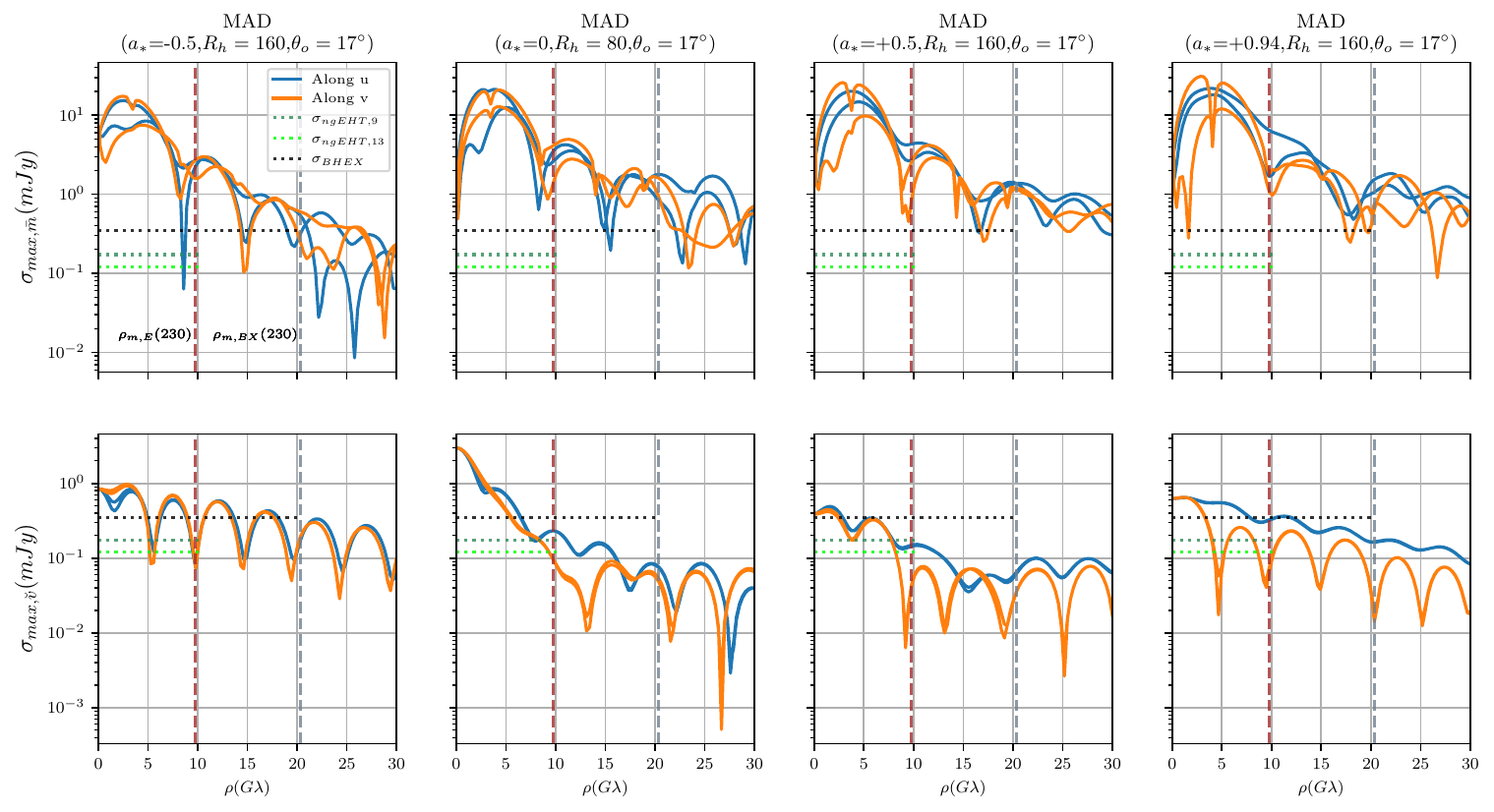}
    \caption{The maximum permitted thermal noise in $\breve{m}$ and $\breve{v}$ for the four best-models for M87* at 230 GHz. The plots are obtained by sampling the Fourier domain signals of the models along the $u$ and $v$ axes. For both, $\sigma_{\rm ngEHT}$ and $\sigma_{\rm BHEX}$, the reference baseline is formed with ALMA in conjunction with the new site/orbiter parameters given in Table \ref{t:newsite}. The System Equivalent Flux Density (SEFD) of ALMA is assumed to be 74 Jy \citep{EHT_M87_II}. The horizontal dashed lines represent the thermal noise for the ngEHT/BHEX-ALMA baseline, with the subscript $9$ and $13$ signifying 9m and 13m antenna diameters respectively. These lines are terminated at the vertical dashed lines which represent the maximum baselines achievable with the corresponding mission at 230 GHz.}
        \label{f:tnoise_mbrv_vbrv}
\end{figure*}

For the four best-bet models observed at a fixed inclination of $\theta_{0}=17^{\circ}$, the maximum thermal noise requirements are plotted in Figure \ref{f:tnoise_mbrv_vbrv}. Here we consider baselines between a new antenna and ALMA, since the latter is the most sensitive site in the existing EHT array \citep{EHT_M87_II} and has played a crucial role in studying the polarimetric properties of M87* \citep{Goddi:2021}. The reference instrumentation specifications for a potential ngEHT site and the BHEX orbiter are tabulated in Table \ref{t:newsite}. 

\begin{table*}
\begin{tabular}{ |c| c| c| c| c| c| c| c| c| c| } 
\hline
Station Type & $T_{r}$(K) & $T_{\text{atm}}$(K) &$\tau$& $T_{\text{sys}}(K)$ & $\eta_{A}$ & $\Delta$v (GHz) & $\Delta$t (s) & $\eta_{Q}$ \\
\hline
ngEHT Site & 50 & 270 & 0.25 & 110 & 0.8 & 16 & 600 & 0.88 \\
\hline
BHEX Orbiter & 50 & - & - & 50 & 0.8 & 16 & 600 & 0.75 \\
\hline
\end{tabular}
\caption{The putative instrumentation parameters for new ngEHT and BHEX stations. Here, since BHEX is an Earth-space mission, its system temperature is not affected the atmospheric opacity and so can be approximated to be equal to the receiver temperature.}
\label{t:newsite}
\end{table*}

From the sensitivity plots, it is noted that for LP, BHEX is sensitive enough to observe the signal with significant improvements in the observation cadence at baselines longer than Earth's diameter. For the ngEHT, due to limitations of baseline lengths on Earth, LP transitions from the photon ring are likely not detectable despite sufficient sensitivity. For CP, however, the higher sensitivity available on the ground is required. Here, the ngEHT can begin to ``sense" the photon ring, albeit requiring maximum thermal noise to be $<1$mJy, whereas BHEX will not be sensitive to these signals across any baseline lengths.

\subsubsection{Antenna Diameter}
For a baseline between a reference station 1 and a new ngEHT/BHEX antenna 2, the antenna diameter $D_{2}$ for the latter is inversely proportional to the thermal noise $\sigma_{1,2}$ on the baseline. The exact expression is derived in the Appendix \ref{s:antenna_diam}. In the context of our analysis, the plots for antenna diameter requirements are given in Figures \ref{f:diameters_earth} and \ref{f:diameters_space} for the ngEHT site-ALMA and BHEX-ALMA baselines respectively. The requirements for the LP and CP signals are different since for them, $\sigma_{1,2}$ is $\sigma_{\text{max},{\breve{m}}}$ and $\sigma_{\text{max},{\breve{v}}}$ respectively. The detectability inferences from Figures \ref{f:diameters_earth} and \ref{f:diameters_space} are similar to those obtained from studying the thermal noise requirements due to the inverse relationship between the two. However we note that for the ngEHT in probing CP signals, there are clear advantages for choosing a 13 m antenna diameter over 9 m. 

\section{Conclusions and Outlook} \label{sec:conclusions}
In this paper, we have introduced two approaches to obtain the transition radii in the Fourier domain beyond which the photon ring signal begins to dominate over the direct image. Firstly, by geometrical modelling of the $n=0$ and $n=1$ signals as  symmetric, concentric Gaussian rings of equal radii, exact analytical expressions for the transition radii have been found for total intensity, LP and CP. The corresponding formulae are functions solely of tunable model parameters and have a similar mathematical structure. Secondly, a general diagnostic tool, namely CAA, is introduced to systematically obtain the transition radii from the suite of time-averaged GRMHD images. This is applied to the simulation library of M87* at 230 GHz and its polarimetric best-bet models are treated in detail. Lastly, by studying the reference sensitivity and antenna diameter specifications for the ngEHT and BHEX missions, we've explored the extent to which these missions can observe the LP and CP photon ring signatures for these best-bet models.

In the context of the M87* model space probed in this paper, we find that Earth-based photon ring detection may have its strongest prospects in sensitivity-limited, astrophysics-forward CP signatures rather than high-resolution imaging. These signatures are relatively insensitive to morphology but may sense the presence of the photon ring in the near future. For BHEX, the prospects for probing the CP signal are sub-optimal, but an overwhelming majority of the signatures in LP are accessible and the nominal sensitivity of BHEX is sufficiently high that Stokes $I$ and $P$ signatures will likely provide strong morphological sensitivity to the photon ring. However, in the case of morphology that disobeys the trends exploited in this work (for example, if Stokes $V$ flips sign across the direct image, as has been seen in some simulations \citep{Tsunetoe_2021}), large discrepancies from rotational symmetry should be apparent even in the Fourier domain. In such cases, there is no substitute for polarimetric imaging.

A natural extension of this work is to performing a similar analysis of M87* at 345 GHz, possibly with the inclusion of more sophisticated geometric models for the image that are currently being used by the EHT \citep{EHT:2019}. It is also timely to perform broad GRMHD library analyses for photon ring detection prospects for Sgr A* as well. Due to its smaller mass, the dynamical timescales of Sgr A* are much shorted than M87* \citep{EHT_SgrA_V} and so it would be instructive to apply the CAA scheme developed in this paper to quantify the range of transition radii as the source evolves. While there has been recent encouraging results in studies of LP detections of the \s{} photon ring with the ngEHT \citep{Shavelle:2024}, BHEX may face challenges in detecting \s{} due to the diminishment of long-baseline signals caused by diffractive scattering \citep{Johnson_Narayan_2016}. Appropriate orbit selection and observation strategies for the Galactic Center are the object of ongoing work; however, interferometric quotients such as those studied in this paper are innately robust to convolutional corruptions to the image, but will be sensitivity-limited in applications to \s{}.

\section{Acknowledgements}
We thank Paul Tiede for his assistance in the application of \texttt{VIDA}, as well as Michael Johnson and Sheperd Doeleman for many helpful conversations. A.T. gratefully acknowledges the Center for Astrophysics $\vert$ Harvard and Smithsonian, and the Black Hole Initiative for several fruitful discussions and providing a welcoming environment during an extended research visit during May-July 2024. 
We acknowledge financial support from the National Science Foundation (AST-2307887). This project was funded in part by generous support from Mr. Michael Tuteur and Amy Tuteur, MD. 
This work was supported by the Black Hole Initiative, which is funded by grants from the John Templeton Foundation (Grant 62286) and the Gordon and Betty Moore Foundation (Grant GBMF-8273) - although the opinions expressed in this work are those of the author(s) and do not necessarily reflect the views of these Foundations.

%



\clearpage
\appendix
\section{Image and Visibility Domain Formulation of the Model}\label{sec:model_theory}
\subsection{The Image Domain}
Consider that the sky image of the black hole is constructed as a convolution between an infinitesimally thin ring of diameter $d$ convolved with a Gaussian of width $\sigma$.
Suppose the image plane is spanned by polar co-ordinates $(\rho_{i},\phi)$ of a thin ring of radius $r_{0}$, having peak intensity $\mathcal{I}_{0}$ is given by a delta function \citep{Johnson_2020}
\begin{gather}
    I_{\delta}(\rho_{i},\phi)\equiv I(\rho_{i})=\frac{\mathcal{I}_{0}}{2\pi r_{0}}\delta\big(\rho_{i}-r_{0}\big). \label{eq:delta_ring}
\end{gather}
Here $\mathcal{I}_{0}$ is constructed such that $2\pi r_{0} \mathcal{I}_{0}$ is the total flux density of the image. The Gaussian function having width $\sigma$ can be modelled as:
\begin{gather}
    I_{G}(\rho_{i},\sigma)=\frac{1}{2\pi\sigma^{2}}\text{exp}\bigg(-\frac{\rho^{2}_{i}}{2\sigma^{2}}\bigg). \label{eq:gaussian_ring}
\end{gather}
Several ring modelling papers model the Gaussian in terms of the FWHM \citep{EHT_M87_IV,Tiede:2022}.

Note that due to the absence of any $\phi$ dependence in the equations above, the model is axisymmetric. Now, to construct the geometric model for the image, we require a convolution of the signals given in Equations \ref{eq:delta_ring} and \ref{eq:gaussian_ring}. The formulation in the image plane is given by \citep{Avendano:2023}:
\begin{gather}
    I(\rho_{i};\sigma)=\int_{0}^{2\pi}\int_{0}^{\infty}I_{\delta}(\rho')I_{G}(\rho_{i}^{2}+\rho'^{2}-2\rho\rho'\cos\phi,\sigma)\rho' d\rho' d\phi \label{eq:convol}.
\end{gather}
Substituting Equation \ref{eq:delta_ring} and Equation \ref{eq:gaussian_ring} into Equation \ref{eq:convol}, we get
\begin{gather}
    I(\rho_{i};\sigma)=\frac{\mathcal{I}_{0}}{(2\pi r_{0})(2\pi\sigma^{2})}\int_{0}^{2\pi}\int_{0}^{\infty}\delta(\rho'-r_{0})\text{exp}\bigg(-\frac{[\rho^{2}_{i}+\rho'^{2}-2\rho\rho'\cos\phi]}{2\sigma^{2}}\bigg)\rho'd\rho'd\phi.
\end{gather}
Separating the integral over $\phi$ gives
\begin{gather}
    I(\rho_{i};\sigma)=\frac{\mathcal{I}_{0}}{(2\pi)^{2}r_{0}\sigma^{2}}\int_{0}^{2\pi}\text{exp}\bigg(\frac{\rho\rho'\cos\phi}{\sigma^{2}}\bigg)d\phi\int_{0}^{\infty}\delta(\rho'-r_{0})\text{exp}\bigg(-\frac{[\rho^{2}_{i}+\rho'^{2}]}{2\sigma^{2}}\bigg)\rho'd\rho'. \label{eq:convol_1}
\end{gather}
Now, the $\phi$ integral can be written in terms of the modified Bessel function of the zeroth order
\begin{gather}
    I_{0}(x)=\frac{1}{\pi}\int_{0}^{\pi}\text{exp}(x\cos\theta)d\theta =\frac{1}{2\pi}\int_{0}^{2\pi}\text{exp}(x\cos\theta)d\theta,
\end{gather}
with the factor of two 2 arising from the fact that the integrand is even. Thus, the Equation \ref{eq:convol_1} now becomes
\begin{gather}
    I(\rho_{i};\sigma)=\frac{\mathcal{I}_{0}}{(2\pi)^{2}r_{0}\sigma^{2}}\int_{0}^{\infty}\Bigg(2\pi I_{0}\bigg(\frac{\rho_{i}\rho'}{\sigma^{2}}\bigg)\Bigg)\delta(\rho'-r_{0})\text{exp}\bigg(-\frac{[\rho^{2}_{i}+\rho'^{2}]}{2\sigma^{2}}\bigg)\rho'd\rho',\nonumber\\
    =\frac{\mathcal{I}_{0}}{2\pi r_{0}\sigma^{2}}\Bigg(I_{0}\bigg(\frac{\rho_{i} r_{0}}{\sigma^{2}}\bigg)\text{exp}\bigg(-\frac{[\rho^{2}_{i}+r_{0}^{2}]}{2\sigma^{2}}\bigg)r_{0}\Bigg), \nonumber\\
    =\frac{\mathcal{I}_{0}}{2\pi\sigma^{2}}I_{0}\bigg(\frac{\rho_{i} r_{0}}{\sigma^{2}}\bigg)\exp\bigg(-\frac{[\rho^{2}_{i}+r_{0}^{2}]}{2\sigma^{2}}\bigg).
\end{gather}
This is the image-domain formulation of a thin ring of radius $r_{0}$ having a Gaussian width $\sigma$.

\subsection{The Visibility Domain}
Now for the delta function in Equation \ref{eq:convol_1}, it is known that the two-dimensional Fourier transform is given by the Bessel function of the first kind of zeroth order $J_{0}(z)$ (upto a scale factor $\mathcal{I}_{0})$),
\begin{gather}
    \tilde{V}_{\delta}(\rho,\psi)=\mathcal{I}_{0}J_{0}(2\pi r_{0}\rho).
\end{gather}
This can be inferred from the fact that for axisymmetric image models, the two dimensional Fourier transform is simply a Hankel transform and the latter for a delta function is simply the Bessel function of zeroth order of the first kind \citep{Thompson}.

Furthermore, it is known that the Fourier transform of a Gaussian is also a Gaussian and so for a function of kind in Equation \ref{eq:gaussian_ring}, the visibility domain signature is:
\begin{gather}
    \tilde{V}_{G}(\rho,\psi)=\text{exp}(-2\pi^{2}\sigma^{2}\rho^{2}).
\end{gather}
Therefore, since the Fourier transforms of a convolution of two functions is simply the product of the Fourier transform of each of the functions, the signature of our ring model in the visibility domain is:
\begin{gather}
    \tilde{V}(\rho,\psi)\equiv \tilde{V}(\rho,\sigma,r_{0})=\tilde{V}_{\delta}V_{G}=\mathcal{I}_{0}J_{0}(2\pi r_{0}\rho)\text{exp}(-2\pi^{2}\sigma^{2}\rho^{2}). \label{eq:visib_gauss_ring}
\end{gather}
This form is common to both $n=0$ and $n=1$ features in our model and by using appropriate scale factors can be mapped to LP and CP models as done in Equations \ref{eq:P01} and \ref{eq:V_01} respectively.
\section{Analytic Expressions for Photon Ring Transition Signatures in Polarimetric Quantities}\label{sec:transition_deriv}
Using the expressions for the Stokes parameters in the Fourier domain for the geometric model given in \ref{eq:I_01} and \ref{eq:P01}, we now derive the transition radii in total intensity and LP. The derivation for CP is given in the main text in Section \ref{sec:ring_with_rho_formulas}.

\subsection{Total Intensity (I)}
For Stokes $I$ in the visibility domain, the dominance of the $n=1$ signature can be intuitively understood with the ratio of the $n=1$ to the $n=0$ term being greater than one. From Equation \ref{eq:I_01}, this implies,
\begin{gather}
\frac{F_{1}J_{0}(2\pi r_{1}\rho)e^{-2\pi^{2}\sigma_{1}^{2}\rho^{2}}}{F_{0}J_{0}(2\pi r_{0}\rho)e^{-2\pi^{2}\sigma_{0}^{2}\rho^{2}}}=\Bigg(\frac{F_{1}}{F_{0}}\Bigg)\Bigg(\frac{J_{0}(2\pi r_{1}\rho)}{J_{0}(2\pi r_{0}\rho)}\Bigg)e^{2\pi^{2}(\sigma_{0}^{2}-\sigma_{1}^{2})\rho^{2}}>1.
\end{gather}
Physically, this relative dominance is expected to occur at longer baselines \citep{Johnson_2020}. Now, assuming equal radii, the $J_{0}(..)$ terms cancel. Then, making the right equal to 1 to obtain the critical/transition value of $\rho$, $\rho_{T}$, we get
\begin{gather}
    e^{2\pi^{2}(\sigma_{0}^{2}-\sigma_{1}^{2})\rho_{T}^{2}}=\frac{F_{0}}{F_{1}}.
\end{gather}
Taking the natural logarithm of both sides, after some straightforward algebra, one obtains the expression for the transition radius in total intensity:
\begin{gather}
\rho_{T}\rightarrow(\rho_{T})_{I}=\sqrt{\frac{\ln\Big(\frac{F_{0}}{F_{1}}\Big)}{2\pi^{2}(\sigma_{0}^{2}-\sigma_{1}^{2})}} \label{eq:rho_t_I}.
\end{gather}

\subsection{Linear polarization}
For LP, the expression for $\breve{m}$ for our ring model, using Equations \ref{eq:P01} and \ref{eq:I_01}, is given by:
\begin{gather}
    \breve{m}_{01}=\frac{\tilde{P}_{01}}{\tilde{I}_{01}}=\frac{-\beta_{2,0}F_{0}e^{2i\theta}J_{2}(2\pi r_{0}\rho)e^{-2\pi^{2}\sigma_{0}^{2}\rho^{2}}-\beta_{2,1}F_{1}e^{2i\theta}J_{2}(2\pi r_{1}\rho)e^{-2\pi^{2}\sigma_{1}^{2}\rho^{2}}}{F_{0}J_{0}(2\pi r_{0}\rho)e^{-2\pi^{2}\sigma_{0}^{2}\rho^{2}}+F_{1}J_{0}(2\pi r_{1}\rho)e^{-2\pi^{2}\sigma_{1}^{2}\rho^{2}}}. \label{eq:mbreve_start}
\end{gather}
Now, we make two approximations. Firstly, we assume equal radii of the $n=0$ and $n=1$ rings, i.e $r_{0}=r_{1}$. Secondly, we assume,
\begin{gather}
J_{2}(2\pi r_{0} \rho)\approx-J_{0}(2\pi r_{0} \rho); \quad \rho \gtrapprox 15G \lambda,
\end{gather}
which is valid in the range of Earth-space baselines of interest to future observations \citep{Palumbo:2023}. Now, as a consequence of this, the Bessel function terms cancel out of the numerator and the denominator, and after cancelling out $F_{0}e^{-2\pi^{2}\sigma_{0}^{2}\rho^{2}}$ we are left with,
\begin{gather}
\bar{m}\equiv\breve{m}_{01}e^{-2i\theta}=\frac{\beta_{2,0}F_{0}e^{-2\pi^{2}\sigma_{0}^{2}\rho^{2}}+\beta_{2,1}F_{1}e^{-2\pi^{2}\sigma_{1}^{2}\rho^{2}}}{F_{0}e^{-2\pi^{2}\sigma_{0}^{2}\rho^{2}}+F_{1}e^{-2\pi^{2}\sigma_{1}^{2}\rho^{2}}}.
\end{gather}
Now, asserting the requirement of criticality as in total intensity, the mathematical condition to be solved for the transition radii $\rho_{T}$ is,
\begin{gather}
    \frac{\vert\beta_{2,1}F_{1}e^{-2\pi^{2}\sigma_{1}^{2}\rho_{T}^{2}}\vert}{\vert\beta_{2,0}F_{0}e^{-2\pi^{2}\sigma_{0}^{2}\rho_{T}^{2}}\vert}=\Bigg(\bigg\lvert\frac{\beta_{2,0}}{\beta_{2,1}}\bigg\rvert\frac{F_{0}}{F_{1}}\Bigg)e^{2\pi^{2}(\sigma_{0}^{2}-\sigma_{1}^{2})\rho_{T}^{2}}=1.
\end{gather}
Here since only $\beta_{2,0}$ and $\beta_{2,1}$ are complex quantities, using the magnitude corresponded to using their absolute values. Taking $\beta_{2,1}F_{1}/\beta_{2,0}F_{0}$ to the right side, performing a natural logarithm on both sides, and solving for $\rho_{T}$, gives:

\begin{gather}
\rho_{T}\rightarrow(\rho_{T})_{\text{LP}}=\sqrt{\frac{\ln\Bigg(\bigg\lvert\frac{\beta_{2,0}}{\beta_{2,1}}\bigg\rvert\frac{F_{0}}{F_{1}}\Bigg)}{2\pi^{2}(\sigma_{0}^{2}-\sigma_{1}^{2})}}. \label{eq:rho_t_mbreve}
\end{gather}

\section{Transition Radii for M87* at 230 GHz}\label{sec:grmhd_geom_table}
The Table  lists the transition radii in LP and CP obtained using the Equations \ref{eq:lp_trans} and \ref{eq:cp_trans} as well as from time-averaged GRMHD simulations of M87* at 230 GHz. The rows in bold are the best-bet models for M87* hedged by the recent polarimetric results of the EHT for M87* \citep{EHT_M87_IX}. 
\clearpage

\begin{center}\label{t:grmhd_geometric_vals}
    \begin{longtable}
    {|c|c|c|c|c|c|c|c|c|c|c|c|c|c|c|c|}   
    \caption{The transition radii $\rho$ in LP and CP for M87* at 230 GHz, denoted by the subscripts $L$ and $C$ respectively. The superscripts $G$ and $S$ denote values obtained from geometric formulae and directly from time-averaged images respectively. The symbol $\mathcal{S}(..)$ denotes the sign of the corresponding quantity.}\\
        \hline
    Flux & $a_{*}$ & $R_{\text{h}}$ & $\sigma_{0}$ & $\sigma_{1}$ & $\mathcal{F}$ &$\mathcal{S}(\angle\beta_{2,0})$ &$\mathcal{S}(\angle\beta_{2,1})$ &$\mathcal{S}(v_{0})$ & $\mathcal{S}(v_{1})$ & $|\beta_{r}|$ & $|v_{r}|$& $\rho_{L}^{G}$ & $\rho_{L}^{S}$ & $\rho_{C}^{G}$ & $\rho_{C}^{S}$\\
    \hline
MAD & -0.94 & 1 & 7.17 & 0.67 & 5.6 & -1.0 & 1.0 & 1.0 & -1.0 & 1.03 & 1.55 & 8.61 & 12.44 & 9.56 & 1.06 \\
MAD & -0.94 & 10 & 6.92 & 0.66 & 5.15 & -1.0 & 1.0 & 1.0 & -1.0 & 1.23 & 0.62 & 9.17 & 13.18 & 7.29 & 1.16 \\
MAD & -0.94 & 20 & 6.91 & 0.66 & 4.96 & -1.0 & 1.0 & 1.0 & -1.0 & 1.37 & 0.49 & 9.35 & 13.69 & 6.33 & 1.04 \\
MAD & -0.94 & 40 & 6.75 & 0.66 & 4.75 & -1.0 & 1.0 & 1.0 & -1.0 & 1.8 & 0.45 & 10.12 & 13.83 & 6.07 & 1.22 \\
MAD & -0.94 & 80 & 6.39 & 0.64 & 4.52 & -1.0 & 1.0 & 1.0 & -1.0 & 2.58 & 0.52 & 11.44 & 13.37 & 6.71 & 4.89 \\
MAD & -0.94 & 160 & 5.9 & 0.61 & 4.25 & -1.0 & 1.0 & 1.0 & -1.0 & 4.05 & 0.68 & 13.35 & 13.06 & 8.13 & 5.94 \\
MAD & -0.5 & 1 & 6.44 & 0.46 & 5.58 & -1.0 & -1.0 & 1.0 & 1.0 & 0.81 & 2.88 & 8.88 & 0.0 & 12.05 & 0.0 \\
MAD & -0.5 & 10 & 6.76 & 0.47 & 4.88 & -1.0 & 1.0 & 1.0 & -1.0 & 0.99 & 2.14 & 8.63 & 5.81 & 10.53 & 6.04 \\
MAD & -0.5 & 20 & 6.98 & 0.48 & 4.65 & -1.0 & 1.0 & 1.0 & -1.0 & 1.12 & 0.84 & 8.55 & 5.38 & 7.76 & 5.77 \\
MAD & -0.5 & 40 & 6.97 & 0.48 & 4.4 & -1.0 & 1.0 & 1.0 & -1.0 & 1.57 & 0.5 & 9.29 & 5.41 & 5.95 & 5.08 \\
MAD & -0.5 & 80 & 6.69 & 0.47 & 4.13 & -1.0 & 1.0 & 1.0 & -1.0 & 2.65 & 0.41 & 10.76 & 5.95 & 5.05 & 4.79 \\
\textbf{MAD} & \textbf{-0.5} & \textbf{160} & \textbf{6.26} & \textbf{0.46} & \textbf{3.86} & \textbf{-1.0} & \textbf{1.0}. & \textbf{1.0} & \textbf{-1.0} & \textbf{6.36} & \textbf{0.4} & \textbf{13.3} & \textbf{11.01} & \textbf{4.91} & \textbf{4.87} \\
MAD & 0 & 1 & 5.73 & 0.47 & 5.8 & -1.0 & 1.0 & -1.0 & -1.0 & 0.98 & 1.37 & 10.72 & 15.02 & 11.71 & 0.0 \\
MAD & 0 & 10 & 5.83 & 0.45 & 4.98 & -1.0 & 1.0 & -1.0 & -1.0 & 1.0 & 0.74 & 10.12 & 14.41 & 9.11 & 0.0 \\
MAD & 0 & 20 & 6.22 & 0.46 & 4.82 & -1.0 & 1.0 & -1.0 & -1.0 & 1.0 & 0.73 & 9.38 & 15.72 & 8.37 & 0.0 \\
MAD & 0 & 40 & 6.63 & 0.46 & 4.68 & -1.0 & 1.0 & -1.0 & -1.0 & 1.19 & 0.8 & 9.21 & 0.0 & 8.06 & 0.0 \\
\textbf{MAD} & \textbf{0} & \textbf{80} & \textbf{7.0} & \textbf{0.46} & \textbf{4.52} & \textbf{-1.0} & \textbf{1.0} & \textbf{-1.0} & \textbf{-1.0} & \textbf{2.18} & \textbf{1.19} & \textbf{10.06}& \textbf{0.0} & \textbf{8.63} & \textbf{0.0} \\
MAD & 0 & 160 & 7.29 & 0.46 & 4.34 & -1.0 & 1.0 & -1.0 & -1.0 & 4.06 & 1.49 & 10.81 & 0.0 & 8.72 & 0.0 \\
MAD & +0.5 & 1 & 5.17 & 0.48 & 5.6 & -1.0 & 1.0 & -1.0 & 1.0 & 0.99 & 0.34 & 11.81 & 15.98 & 7.19 & 5.22 \\
MAD & +0.5 & 10 & 4.96 & 0.47 & 4.85 & -1.0 & 1.0 & -1.0 & 1.0 & 1.02 & 0.35 & 11.91 & 13.71 & 6.8 & 5.63 \\
MAD & +0.5 & 20 & 5.12 & 0.47 & 4.77 & -1.0 & 1.0 & -1.0 & 1.0 & 1.08 & 0.41 & 11.65 & 13.43 & 7.51 & 6.83 \\
MAD & +0.5 & 40 & 5.24 & 0.48 & 4.73 & -1.0 & 1.0 & -1.0 & 1.0 & 1.47 & 0.62 & 12.38 & 12.73 & 9.25 & 7.45 \\
MAD & +0.5 & 80 & 5.32 & 0.48 & 4.7 & -1.0 & 1.0 & -1.0 & 1.0 & 2.74 & 6.51 & 14.02 & 12.75 & 16.22 & 4.92 \\
\textbf{MAD} & \textbf{+0.5} & \textbf{160} & \textbf{5.38} & \textbf{0.48} & \textbf{4.66} & \textbf{-1.0} & \textbf{1.0} & \textbf{1.0} & \textbf{-1.0} & \textbf{4.67} & \textbf{0.43} & \textbf{15.21} & \textbf{10.34} & \textbf{7.18} & \textbf{20.69} \\
MAD & +0.94 & 1 & 5.73 & 0.7 & 5.54 & -1.0 & 1.0 & -1.0 & 1.0 & 1.01 & 0.28 & 10.71 & 15.02 & 5.38 & 0.0 \\
MAD & +0.94 & 10 & 5.35 & 0.67 & 5.01 & -1.0 & 1.0 & -1.0 & 1.0 & 1.02 & 0.12 & 11.18 & 14.43 & nan & 0.0 \\
MAD & +0.94 & 20 & 5.42 & 0.68 & 4.99 & -1.0 & 1.0 & -1.0 & 1.0 & 1.04 & 0.1 & 11.09 & 14.45 & nan & 0.0 \\
MAD & +0.94 & 40 & 5.52 & 0.68 & 5.04 & -1.0 & 1.0 & -1.0 & 1.0 & 1.19 & 0.09 & 11.36 & 13.17 & nan & 0.0 \\
MAD & +0.94 & 80 & 5.57 & 0.67 & 5.12 & -1.0 & 1.0 & -1.0 & 1.0 & 1.72 & 0.07 & 12.39 & 9.6 & nan & 0.0 \\
\textbf{MAD} & \textbf{+0.94} & \textbf{160} & \textbf{5.58} & \textbf{0.66} & \textbf{5.21} & \textbf{-1.0} & \textbf{1.0} & \textbf{-1.0} & \textbf{1.0} & \textbf{3.17} & \textbf{0.02} & \textbf{14.02} & \textbf{9.6} & \textbf{nan} & \textbf{0.0} \\
SANE & -0.94 & 1 & 11.52 & 0.94 & 13.96 & -1.0 & 1.0 & 1.0 & 1.0 & 5.46 & 0.72 & 8.42 & 16.03 & 6.14 & 0.0 \\
SANE & -0.94 & 10 & 6.95 & 0.61 & 5.86 & -1.0 & 1.0 & 1.0 & -1.0 & 32.14 & 1.33 & 15.36 & 6.03 & 9.62 & 11.03 \\
SANE & -0.94 & 20 & 6.57 & 0.58 & 5.67 & -1.0 & 1.0 & 1.0 & -1.0 & 38.0 & 0.71 & 16.45 & 6.17 & 8.37 & 6.0 \\
SANE & -0.94 & 40 & 6.28 & 0.56 & 5.47 & -1.0 & 1.0 & 1.0 & -1.0 & 36.26 & 0.56 & 17.08 & 6.8 & 7.87 & 6.07 \\
SANE & -0.94 & 80 & 6.03 & 0.54 & 5.28 & -1.0 & 1.0 & 1.0 & -1.0 & 32.35 & 0.52 & 17.52 & 12.41 & 7.73 & 6.67 \\
SANE & -0.94 & 160 & 5.85 & 0.53 & 5.11 & -1.0 & 1.0 & 1.0 & -1.0 & 29.82 & 0.54 & 17.87 & 12.65 & 8.05 & 11.12 \\
SANE & -0.5 & 1 & 10.29 & 0.58 & 17.88 & -1.0 & 1.0 & 1.0 & 1.0 & 0.93 & 0.7 & 7.57 & 4.31 & 7.19 & 0.0 \\
SANE & -0.5 & 10 & 7.11 & 0.58 & 5.3 & -1.0 & 1.0 & 1.0 & 1.0 & 9.25 & 1.26 & 12.92 & 8.39 & 9.03 & 0.0 \\
SANE & -0.5 & 20 & 6.91 & 0.42 & 5.05 & -1.0 & 1.0 & 1.0 & 1.0 & 9.67 & 3.43 & 13.28 & 9.6 & 11.37 & 0.0 \\
SANE & -0.5 & 40 & 6.69 & 0.4 & 4.84 & -1.0 & 1.0 & 1.0 & -1.0 & 9.16 & 9.22 & 13.53 & 9.3 & 13.54 & 10.55 \\
SANE & -0.5 & 80 & 6.45 & 0.39 & 4.56 & -1.0 & 1.0 & 1.0 & -1.0 & 9.51 & 2.26 & 13.99 & 6.4 & 11.01 & 10.53 \\
SANE & -0.5 & 160 & 6.25 & 0.37 & 4.24 & -1.0 & 1.0 & 1.0 & -1.0 & 8.22 & 1.25 & 14.03 & 5.93 & 9.61 & 10.55 \\
SANE & 0 & 1 & 8.24 & 0.49 & 18.48 & 1.0 & -1.0 & -1.0 & -1.0 & 0.56 & 1.1 & 8.62 & 6.59 & 9.8 & 0.0 \\
SANE & 0 & 10 & 9.38 & 0.68 & 10.09 & -1.0 & -1.0 & -1.0 & -1.0 & 2.02 & 1.25 & 8.62 & 0.0 & 7.9 & 0.0 \\
\hline
\newpage
\hline
    Flux & $a$ & $R_{\text{h}}$ & $\sigma_{0}$ & $\sigma_{1}$ & $\mathcal{F}$ &$\mathcal{S}(\angle\beta_{2,0})$ &$\mathcal{S}(\angle\beta_{2,1})$ &$\mathcal{S}(v_{0})$ & $\mathcal{S}(v_{1})$ & $|\beta_{r}|$ & $|v_{r}|$& $\rho_{L}^{G}$ & $\rho_{L}^{S}$ & $\rho_{C}^{G}$ & $\rho_{C}^{S}$\\
    \hline
SANE & 0 & 20 & 6.87 & 0.37 & 5.26 & -1.0 & 1.0 & -1.0 & -1.0 & 10.76 & 1.8 & 13.6 & 5.98 & 10.15 & 0.0 \\
SANE & 0 & 40 & 6.55 & 0.32 & 4.49 & -1.0 & 1.0 & -1.0 & -1.0 & 11.68 & 2.22 & 14.11 & 5.99 & 10.76 & 0.0 \\
SANE & 0 & 80 & 6.38 & 0.3 & 4.31 & -1.0 & 1.0 & -1.0 & -1.0 & 8.42 & 2.83 & 13.8 & 5.98 & 11.52 & 0.0 \\
SANE & 0 & 160 & 6.2 & 0.29 & 4.18 & -1.0 & 1.0 & -1.0 & -1.0 & 7.49 & 2.89 & 13.91 & 6.03 & 11.83 & 0.0 \\
SANE & +0.5 & 1 & 7.39 & 0.65 & 10.21 & -1.0 & 1.0 & -1.0 & 1.0 & 0.92 & 0.23 & 9.45 & 5.78 & 5.81 & 4.65 \\
SANE & +0.5 & 10 & 6.56 & 0.66 & 7.53 & 1.0 & -1.0 & -1.0 & -1.0 & 0.7 & 1.48 & 9.18 & 20.95 & 11.04 & 0.0 \\
SANE & +0.5 & 20 & 5.86 & 0.67 & 6.74 & -1.0 & -1.0 & -1.0 & -1.0 & 0.41 & 1.7 & 8.07 & 0.0 & 12.46 & 0.0 \\
SANE & +0.5 & 40 & 5.71 & 0.71 & 5.56 & -1.0 & -1.0 & -1.0 & -1.0 & 1.35 & 1.91 & 11.64 & 0.0 & 12.6 & 0.0 \\
SANE & +0.5 & 80 & 5.82 & 0.87 & 4.42 & -1.0 & -1.0 & -1.0 & -1.0 & 3.56 & 0.68 & 13.4 & 0.0 & 8.47 & 0.0 \\
SANE & +0.5 & 160 & 5.6 & 1.23 & 3.82 & -1.0 & 1.0 & -1.0 & -1.0 & 5.47 & 0.41 & 14.83 & 4.81 & 5.63 & 0.0 \\
SANE & +0.94 & 1 & 6.12 & 0.77 & 6.91 & -1.0 & 1.0 & 1.0 & 1.0 & 1.06 & 0.01 & 10.79 & 13.8 & nan & 0.0 \\
SANE & +0.94 & 10 & 4.89 & 0.71 & 5.85 & -1.0 & 1.0 & -1.0 & 1.0 & 2.4 & 0.55 & 15.6 & 8.58 & 10.37 & 10.23 \\
SANE & +0.94 & 20 & 5.04 & 0.74 & 5.56 & -1.0 & 1.0 & -1.0 & 1.0 & 5.69 & 1.95 & 17.29 & 9.35 & 14.37 & 11.6 \\
SANE & +0.94 & 40 & 12.06 & 0.79 & 4.51 & -1.0 & 1.0 & -1.0 & 1.0 & 15.37 & 0.73 & 7.94 & 10.15 & 4.2 & 4.07 \\
SANE & +0.94 & 80 & 6.56 & 0.84 & 3.84 & -1.0 & 1.0 & -1.0 & 1.0 & 14.93 & 0.03 & 14.36 & 9.7 & nan & 2.57 \\
SANE & +0.94 & 160 & 7.08 & 0.95 & 3.54 & -1.0 & 1.0 & 1.0 & 1.0 & 17.11 & 0.27 & 13.4 & 9.13 & nan & 0.0 \\
\hline
    \end{longtable}
\end{center}

\section{Sensitivity and Antenna Diameter Considerations} \label{sec:sensitivity_tn}
\subsection{Maximum Permissible Thermal Noise}
Here we derive the expression for the maximum thermal noise in the CP ($\sigma_{\text{max},\breve{v}}$) and LP ($\sigma_{\text{max},\breve{m}}$) signal respectively. 

Starting with CP, since $\breve{v}=\tilde{V}/\tilde{I}$, the error  $\sigma_{\breve{V}}$ is given by, 
\begin{gather}
    \frac{\sigma_{\breve{v}}^{2}}{|\breve{v}|^{2}}=\frac{\sigma_{\tilde{V}}^{2}}{|\tilde{V}|^{2}}+\frac{\sigma_{\tilde{I}}^{2}}{|\tilde{I}|^{2}} \label{eq:cp_error}.
\end{gather}
Assuming $\sigma_{\tilde{V}}=\sigma_{\tilde{I}}\equiv \sigma$, and taking the square root of both sides, we get,
\begin{gather}
    \frac{\sigma_{\breve{v}}}{|\breve{v}|}=\sigma\sqrt{\frac{1}{|\tilde{V}|^{2}}+\frac{1}{|\tilde{I}|^{2}}}.
\end{gather}
Since the left-hand-side is required to be less than 1, to find its maximum value, the right hand-hand-side has to be equated to 1, such that the maximum thermal noise, $\sigma\rightarrow\sigma_{\text{max},\breve{v}}$, in $\breve{v}$ is:
\begin{gather}
    \sigma_{\text{max},\breve{v}}=\frac{1}{\sqrt{\frac{1}{|\tilde{V}|^{2}}+\frac{1}{|\tilde{I}|^{2}}}}.
\end{gather}
For LP, note that $\tilde{P}$ can be written in terms of Stokes $\tilde{Q}$ and $\tilde{U}$ as
\begin{gather}
    \tilde{P}=\tilde{Q}+i\tilde{U},
\end{gather}
and so assuming $\sigma_{\tilde{Q}}\approx \sigma_{\tilde{U}}=\sigma'$, we get:
\begin{gather}
\sigma_{\tilde{P}}^{2}=\sigma_{\tilde{Q}}^{2}+\sigma_{\tilde{U}}^{2}=2\sigma'^{2} \label{eq:lp_error_cond}.
\end{gather}
Now, for $\breve{m}=\tilde{P}/\tilde{I}$, in the high signal-to-noise limit, the error can be approximated in a manner similar to Equation  \ref{eq:cp_error} \citep{Chael:2016}, giving:
\begin{gather}
    \frac{\sigma_{\breve{m}}^{2}}{|\breve{m}|^{2}}=\frac{\sigma_{\tilde{P}}^{2}}{|\tilde{P}|^{2}}+\frac{\sigma_{\tilde{I}}^{2}}{|\tilde{I}|^{2}}. \label{eq:cp_err_eqn}
\end{gather}
Assuming $\sigma'=\sigma$, taking the square root of both sides, using Equation \ref{eq:lp_error_cond} and once again imposing the condition of the right-hand-side to be equal to 1, we get the expression for the maximum thermal noise $\sigma_{\text{max},\breve{m}}$ in $\breve{m}$ as:
\begin{gather}
    \sigma_{\text{max},\breve{m}}=\frac{1}{\sqrt{\frac{2}{|\tilde{P}|^{2}}+\frac{1}{|\tilde{I}|^{2}}}} \label{eq:lp_err_eqn}.
\end{gather}
The Equations \ref{eq:cp_err_eqn} and \ref{eq:lp_err_eqn} are used to generate the plots in Figure \ref{f:tnoise_mbrv_vbrv} in the main text.

\subsection{Quantifying Antenna Diameter Requirements} \label{s:antenna_diam}
\begin{figure}[h]
    \centering
    \includegraphics[width=\columnwidth]{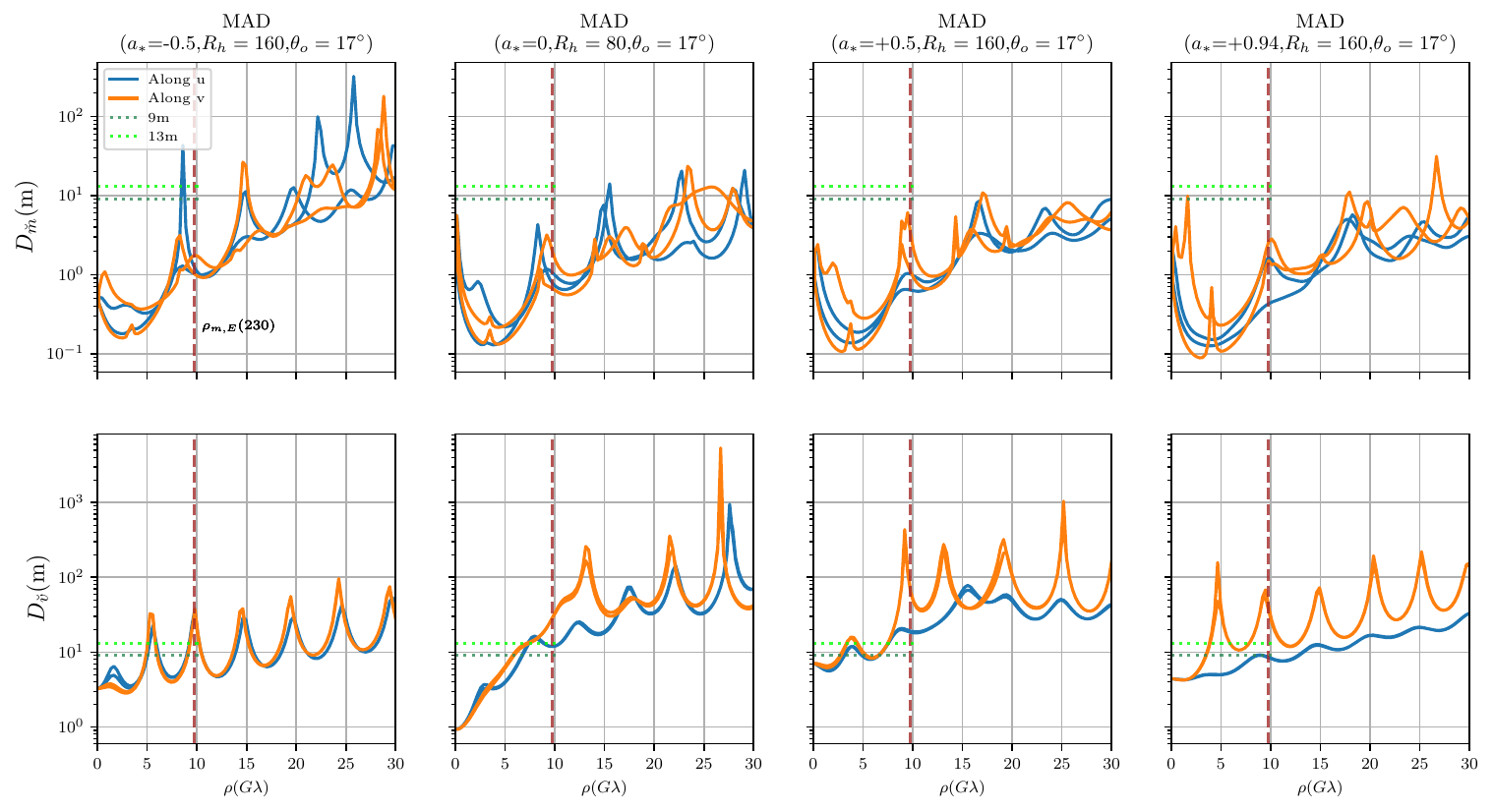}
    \caption{The antenna diameter requirements for an ngEHT site-ALMA baseline. For the site, the putative parameters considered are system temperature $T_{s}=110K$ and antenna efficiency $\eta_{A}=0.8.$ For the baseline, the bandwidth considered is $\Delta v=16$GHz with an integration time of $\Delta t=$600s and (2-bit) quantisation efficiency $\eta_{Q}=$0.88. The SEFD of ALMA is assumed to be 74 Jy.}
        \label{f:diameters_earth}
\end{figure}

Consider the baseline formed by two stations 1 and 2 with the baseline thermal noise given by $\sigma_{1,2}$. This can be written in terms of the station parameters, namely the System Equivalent Flux Densities (SEFD$_{1}$ and SEFD$_{2}$) measured in Jansky (Jy), bandwidth $\Delta v$ measured in Hertz (Hz) and coherence time $\Delta t$ measured in seconds (s),
\begin{gather}
    \sigma_{1,2}(\text{Jy})=\sqrt{\frac{\text{SEFD}_{1}(\text{Jy})\times\text{SEFD}_{2}(\text{Jy})}{2\Delta v(\text{Hz}) \Delta t(\text{s})}} \label{eq:tnoise_def}.
\end{gather}
For any given station, hereafter Station 2, the SEFD is given in terms of the effective system temperature $T_{\text{S},2}$, antenna area $A$ and aperture efficiency $\eta_{A,2}$, by the formula,
\begin{gather}
    \text{SEFD}_{2}(\text{Jy})=\frac{2k_{B}(\text{J}/\text{K})\times T_{S,2}(\text{K})}{\eta_{A,2}\times A(\text{m}^{2})}, \label{eq:sefd_def}
\end{gather}
where $k_{B}=1.380\times10^{-23}$J/K is the Boltzmann constant. Now, since Jansky is \textit{not} a standard (SI) unit while all other quantities are specified in terms of their SI units, it is important to note the following unit conversion:
\begin{gather}
    1\text{Jy}=10^{-26}(\text{J}/\text{m}^{2}) \label{eq:jansky_SI}.
\end{gather}
Keeping this conversion in mind, and approximating for station 2 the area of the single-dish antenna with diameter $D_{2}$ by $\pi D^{2}_{2}/4$, substituting Equation \ref{eq:sefd_def} into Equation \ref{eq:tnoise_def}, making the unit conversion from Equation \ref{eq:jansky_SI} and solving for diameter $D_{2}$, we get:
\begin{gather}
    D_{2}=\frac{1}{\sigma_{1,2}(\text{Jy})\eta_{Q}}\sqrt{\frac{4\times\text{SEFD}_{1}(\text{Jy})\times k_{B}(\text{J/K})\times T_{S,2}(\text{K})\times10^{26}}{\Delta v(\text{Hz}) \times \Delta t(\text{s})\times\eta_{A}\times\pi}}(\text{m}) \label{eq:antenna_diam}.
\end{gather}
This formula is used to compute the plots given in Figure \ref{f:diameters_earth}.

\begin{figure}[h]
    \centering
    \includegraphics[width=\columnwidth]{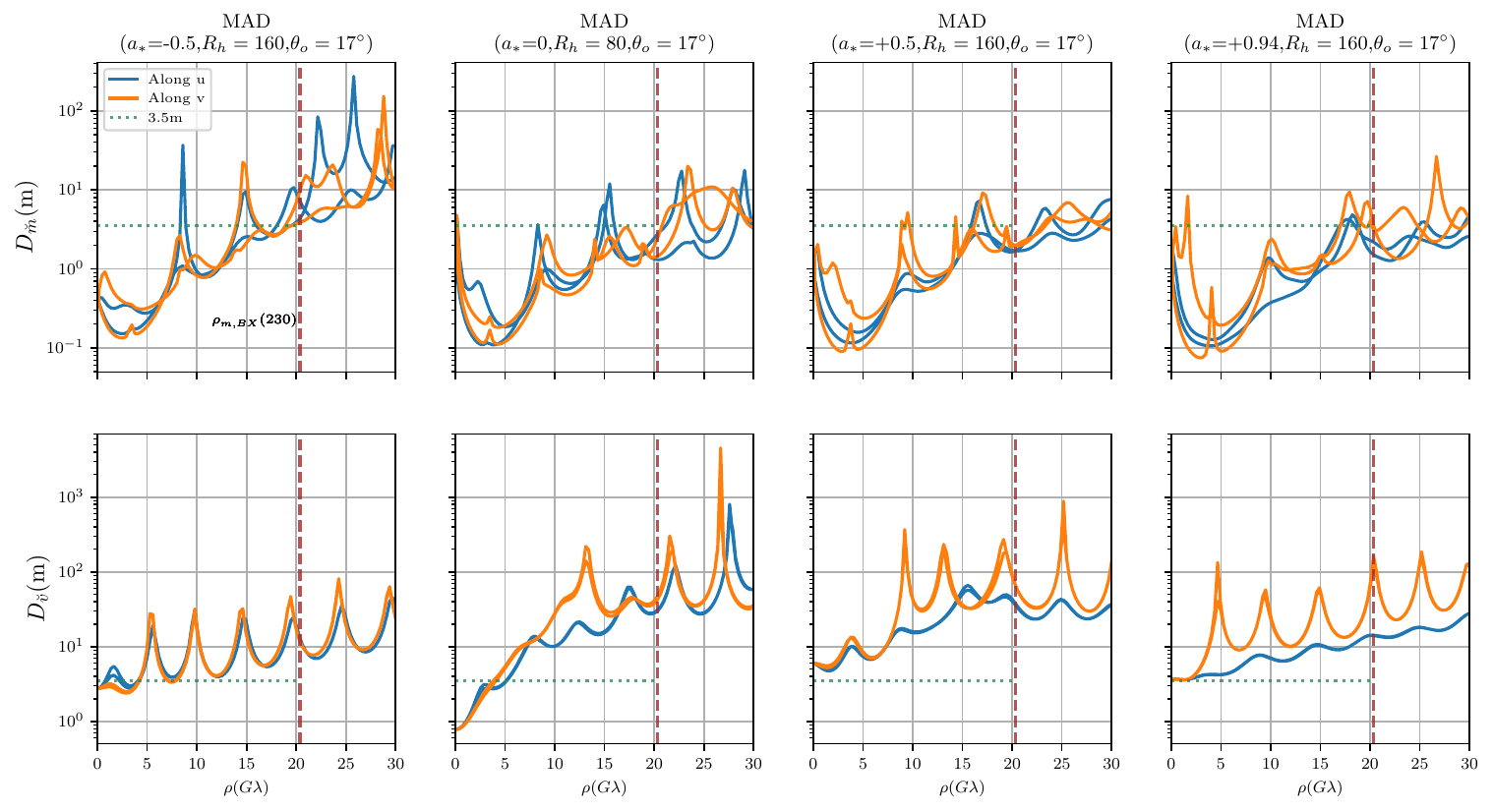}
    \caption{The antenna diameter requirements for the BHEX forming an Earth-Space  baseline with ALMA. The orbiter parameters are given in \ref{t:newsite}.}
        \label{f:diameters_space}
\end{figure}

For the ground-based ngEHT site, a reasonable value of $T_{S,2}$, can be obtained from standard values of the receiver temperature $T_{rx}$ at 230 GHz, forward efficiency of the antenna $\eta_{eff}$, atmospheric temperature $T_{atm}$ and opacity $\tau$ as \citep{Doeleman:2023}, 
\begin{gather}
    T_{S,2}\approx T_{rx}+\eta_{ff}T_{atm}(1-e^{-\tau})=
    50K + 1(270K)(1-e^{-0.25}) \approx 110K.
\end{gather}
For BHEX, the diameter plot is given in Figure \ref{f:diameters_space}. A summary of the parameters used to make the plots is given in Table \ref{t:newsite}.

\clearpage
\bibliography{sample631}{}
\bibliographystyle{aasjournal}



\end{document}